\documentclass[letterpaper]{article} 
\usepackage{aaai2026}
\usepackage{times}  
\usepackage{helvet}  
\usepackage{courier}  
\usepackage[hyphens]{url}  
\usepackage{graphicx} 
\usepackage{natbib}  
\usepackage{caption} 
\usepackage{amsthm}
\usepackage{algorithm}
\usepackage{amsmath, amssymb}
\usepackage{booktabs}
\usepackage{multirow}
\usepackage{xcolor}
\usepackage{float}
\usepackage{afterpage}
\usepackage{algpseudocode}
\newtheorem{lemma}{Lemma}

\usepackage[inline]{enumitem}
\usepackage{longtable}
\usepackage{booktabs}
\usepackage{pifont}
\usepackage{graphicx}
\usepackage{tcolorbox}
\usepackage{multirow}
\usepackage{amsmath,amsthm,amssymb,amsfonts}
\newtheorem{theorem}{Theorem}
\newtheorem{proposition}{Proposition}
\usepackage{amsmath}
\usepackage{float}

\newcommand{\highlight}[1]{{\noindent\textbf{#1}}}
\usepackage{newfloat}
\usepackage{listings}
\DeclareCaptionStyle{ruled}{labelfont=normalfont,labelsep=colon,strut=off} 
\floatstyle{ruled}
\newfloat{listing}{tb}{lst}{}
\floatname{listing}{Listing}
\title{Do Not Merge My Model! Safeguarding Open-Source LLMs Against Unauthorized Model Merging}

\author {
    Qinfeng Li\textsuperscript{\rm 1}\equalcontrib,\hspace{0.4em}
    Miao Pan\textsuperscript{\rm 1}\equalcontrib,\hspace{0.4em}
    Jintao Chen\textsuperscript{\rm 1,4}\thanks{~Corresponding author: \texttt{chenjintao@zju.edu.cn}},\hspace{0.4em}
    Fu Teng\textsuperscript{\rm 1},\\
    Zhiqiang Shen\textsuperscript{\rm 2},\hspace{0.4em}
    Ge Su\textsuperscript{\rm 1},\hspace{0.4em}
    Hao Peng\textsuperscript{\rm 3},\hspace{0.4em}
    Xuhong Zhang\textsuperscript{\rm 1,4}
}

\affiliations {
    \textsuperscript{\rm 1}Zhejiang University\\
    \textsuperscript{\rm 2}Ant Group\\
    \textsuperscript{\rm 3}Zhejiang Normal University\\
    \textsuperscript{\rm 4}Ningbo Global Innovation Center, Zhejiang University\\
}

\usepackage{bibentry}

\begin{document}

\maketitle

\begin{abstract}
Model merging has emerged as an efficient technique for expanding large language models (LLMs) by integrating specialized expert models. However, it also introduces a new threat: model merging stealing, where free-riders exploit models through unauthorized model merging.
Unfortunately, existing defense mechanisms fail to provide effective protection. Specifically, we identify three critical protection properties that existing methods fail to simultaneously satisfy: (1) proactively preventing unauthorized merging; (2) ensuring compatibility with general open-source settings; (3) achieving high security with negligible performance loss.
To address the above issues, we propose MergeBarrier, a plug-and-play defense that proactively prevents unauthorized merging. The core design of MergeBarrier is to disrupt the Linear Mode Connectivity (LMC) between the protected model and its homologous counterparts, thereby eliminating the low-loss path required for effective model merging.
Extensive experiments show that MergeBarrier effectively prevents model merging stealing with negligible accuracy loss.
\end{abstract}

\section{Introduction}

Large language models (LLMs) are widely applied in various fields due to their impressive capabilities~\cite{openai2023gpt4}. However, their knowledge is inherently constrained by their training data~\cite{cong2023have}. For example, Code Llama~\cite{roziere2023code} exhibits limited performance on niche programming languages such as Rust and OCaml~\cite{cassano2024knowledge}.
As a result, expanding LLM capacity has become a major research focus. While fine-tuning with high-quality data—using techniques such as Low-Rank Adaptation~\cite{ijc1.5}—offers a lightweight alternative to full model training, it still demands significant resources. In contrast, model merging~\cite{ijc1.6,ijc1.7} is emerging as a more efficient solution. By integrating multiple homologous expert models specialized in different tasks, a merged model retains their strengths without requiring data collection or high-performance hardware (e.g., GPUs). Given these advantages, model merging is increasingly viewed as a promising direction for enhancing LLMs~\cite{yang2024model}.

However, unauthorized model merging emerges as a new threat that may compromise the intellectual property (IP) rights of open-access proprietary models~\cite{cong2023have,yamabe2024mergeprint}. 
For example, many models hosted on Hugging Face~\cite{huggingface} are released under restrictive open-access licenses like CC BY-NC-ND 4.0, which explicitly prohibit commercial use~\cite{creativecommonsBYNCND4}.
Nevertheless, free riders may use improper means to misuse these models for commercial gain.
The misuse of open-access resources is a well-established issue in the traditional software domain, as exemplified by cases such as Jacobsen v. Katzer~\shortcite{jacobsen2008}.
In the LLM ecosystem, where models possess substantial commercial value, the risk of misuse is further exacerbated by advanced model merging technology. Such techniques allow free riders to incorporate open-access models into their own, making unauthorized appropriation easier. Even worse, existing research indicates that model merging could render watermarks ineffective~\cite{cong2023have}, making such stealing difficult to trace.
We refer to this threat as \textit{model merging stealing}, which is an emerging and serious form of model theft that is easy to perform and difficult to trace.
Given the potential severe impact on the LLM open-source environment, \textit{safeguarding open-access LLMs from unauthorized model merging is a critical issue.}

\begin{table*}[t]
\centering
\resizebox{1.4\columnwidth}{!}{
\begin{tabular}{lccc}
\toprule
Solutions (exemplar)     &  Proactivity  & Compatibility & Security with Utility\\ \midrule

Passive method~\cite{mm1.18,mm2.5}                  & \ding{55} & \checkmark                       & \checkmark     \\
Authorization-based protection~\cite{mm0.0}                & \checkmark   & \ding{55}                    & \checkmark                                                                                            \\
Emulator-based protection~\cite{www4.1}               & \checkmark     & \checkmark                   & \ding{55}                                                                                                 \\
MergeBarrier~(ours)                  & \checkmark   & \checkmark                 & \checkmark                 \\ \bottomrule
\end{tabular}
}
\caption{Comparison with existing solutions. \checkmark/\ding{55} illustrates whether the method can achieve the corresponding property. }
\label{table_intro}
\end{table*}

Unfortunately, as shown in Table \ref{table_intro}, traditional solutions struggle to protect open-source models from model merging stealing as they fail to meet the diverse requirements. First, passive protection methods, such as watermarking~\cite{mm1.18,mm1.19,mm1.20} and fingerprinting~\cite{mm2.5,mm2.6}, only aim to verify model ownership after misuse. Therefore, they are fundamentally incapable of preventing unauthorized merging in the first place. Such passive protection introduces a critical security vulnerability, rendering these methods ineffective: once a model is merged, it may be freely exploited without detection.

In contrast, proactive protections focus on preventing unauthorized usage by enforcing usage constraints, such as limiting model performance when authorization is not provided, thereby mitigating misuse. However, a substantial portion of proactive methods rely on additional components, such as key-based~\cite{mm1.21, mm1.22} or TEE hardware-based~\cite{mm0.0, aaai0.0} authorization systems.
The reliance on these additional authorization components creates compatibility issues in typical open-source environments, where such additional components are not feasible.

Alternatively, a potential method compatible with the open-source scenario is to release an emulator, which replaces the original model, that can perform inference independently~\cite{www4.1,wang2024taylor,junhao2025disrupting}. However, existing emulator-based methods struggle to balance security and utility: simplified emulators reduce utility, while precise ones risk exposing model details. 
For example, in Xiao et al.~\shortcite{www4.1}'s method, the emulator is constructed by dropping layers and using knowledge distillation. While this reduces the risk of exposing the model's inner workings, it also weakens the emulator's performance because many useful model layers are removed. In contrast, TaylorMLP~\cite{wang2024taylor} preserves performance by applying Taylor expansion, but it leaves the model vulnerable by only protecting a single layer in each transformer block, exposing the rest of the model’s weights.

Considering the limitations of existing solutions, we identify three challenges (\textbf{C}) to safeguarding open-source LLMs from model merging stealing. 
\textbf{C1 (Proactivity)}: Achieving proactive protection to ensure the model remains non-mergeable, even when the open-sourced model is fully accessible to the attacker.
\textbf{C2 (Compatibility)}: Achieving protection even when no additional external components can be utilized.
\textbf{C3 (Security with Utility)}: Securing weights sufficiently even without compromising model utility.

To protect open-source LLMs from unauthorized merging, we propose MergeBarrier, a defense that proactively prevents model merging. Our method is based on the insight that model merging relies on Linear Mode Connectivity (LMC) between homologous models~\cite{ainsworth2022git, crisostomi2024c}. Specifically, homologous models are fine-tuned from the same base model. As a result, these models are located within a shared low-loss basin in weight space. Therefore, a linear path exists between these models, along which all intermediate points—representing merged models—also remain within the low-loss basin. This ensures that the merged models maintain high performance.
Based on this insight, to address \textbf{C1}, MergeBarrier disrupts the homogeneity between the protected model and other homologous models. We achieve this by displacing the merged model out of the low-loss basin, thereby preventing merging.
This displacement is achieved by transforming the weights of the protected model into an equivalent set of weights, requiring no external runtime support (address \textbf{C2}).

To address \textbf{C3}, MergeBarrier ensures the performance-preserving transformations are applied throughout the entire transformer block. Specifically, we apply transformations to both the attention and FFN blocks, tailored to their distinct structures.
For the attention block, we leverage its paired linear structure (i.e., query and key linear layers) and apply a shared orthogonal projection to both. This projection is optimized to maximize the deviation between the merged weights and the original weights, thereby displacing the merged model out of the low-loss basin.
Meanwhile, due to the properties of the orthogonal matrix, this projection cancels out during paired linear computation, ensuring that the model's functionality remains unaffected. 
For FFN, we adopt an activation function expansion-based weight reparameterization that rewrites the FFN computation using a new set of polynomial parameters. This hides the original weights, thus disrupting merging. By selecting the optimal expansion point based on the training set, we ensure that the approximation error has a provably negligible impact on model performance. Interestingly, our analysis reveals that this error can serve as a built-in safeguard by reducing weight inversion to an Learning With Errors (LWE) problem~\cite{regev2009lattices}, i.e., a NP-hard problem, which makes such attacks computationally infeasible.

We evaluate MergeBarrier by protecting expert models across different domains and assessing the performance of merged models in the respective tasks.
The evaluation shows that MergeBarrier offers stronger security than existing methods. Once attackers attempt to merge a MergeBarrier-protected model, the resulting merged model suffers a significant performance drop. Besides, MergeBarrier protection does not compromise model accuracy. The contributions of this work are as follows:
\begin{itemize}[leftmargin=10pt]
\item We are the first work that formally defines the problem of defending against \textit{model merging stealing} and systematically identifies the key requirements for protecting open-source models from unauthorized merging.
\item We introduce MergeBarrier, a plug-and-play solution that protects model weights from being merged by applying orthogonal projection to attention layers and activation function expansion to FFN layers, effectively disrupting the merging path while fully preserving the original model’s performance.
\item We make two foundational contributions to model parameter protection: (1) a broadly applicable projection-based transformation for securing model weights; and (2) a theoretical analysis showing that the noise introduced by Taylor expansion makes weight inversion computationally infeasible (NP-hard), thereby establishing strong security guarantees for broader applications of such reparameterization methods.
\item Extensive experiments demonstrate that MergeBarrier provides a stronger security guarantee compared to existing solutions with negligible accuracy loss. 
Importantly, we also analyze the underlying reason for its effectiveness in preventing model merging.
\end{itemize}

\section{Preliminaries}


\subsection{Model Merging}
Model merging refers to the technique of combining the weights of multiple specialized models to create a single unified model. 
A fundamental concept in model merging is the existence of Linear Mode Connectivity (LMC)~\cite{ainsworth2022git}. Specifically, when models are fine-tuned from the same base model, they exhibit certain homologous properties, meaning they share a similar structure and fall within the same low-loss region of the weight space. This homogeneity is crucial for model merging because it ensures that the models are close enough in weight space to be smoothly interpolated without significant loss of performance.
Currently, the most promising model merging methods, including Task Arithmetic~\cite{ilharco2022editing}, TIES-MERGING~\cite{yadav2023ties}, and DARE~\cite{cong2023have}, are detailed in Appendix~\ref{model merging techniques}.

\subsection{Related work.} 
\subsubsection{Passive Methods.}
Passive protection techniques, such as watermarking~\cite{mm1.18, mm1.19, mm1.20} and fingerprinting~\cite{mm2.5, mm2.6}, detect misuse after it occurs. Watermarking embeds unique signals in model parameters for ownership verification, while fingerprinting alters the model’s behavior to track its origin. However, these methods cannot prevent unauthorized merging, leaving the model vulnerable once merged, as ownership may be obscured.

\highlight{Authorization-based Protection.}
Authorization-based methods, like key-based~\cite{mm1.21, mm1.22} and TEE-based~\cite{mm0.0, aaai0.0} protections, restrict access through cryptographic keys or secure hardware. Key-based systems control model usage via authorized keys, but key management is challenging in open-source environments. TEE-based systems ensure execution within trusted environments, providing robust security, but require specialized hardware, making them incompatible with open-source practices.

\highlight{Emulator-based Protection.}
Existing methods struggle to balance security and utility: simplified emulators reduce performance, while precise ones risk leaking model details. Xiao et al.\cite{www4.1} propose an emulator using layer dropping and knowledge distillation, which protects the model but lowers its utility and requires extensive retraining. To improve utility, TaylorMLP\cite{wang2024taylor} uses a Taylor expansion for lossless weight protection, but only safeguards a single linear layer in each transformer block, leaving the model vulnerable. PaRaMS~\cite{junhao2025disrupting} enhances security by permuting MLP weights and scaling attention weights, yet in model merging scenarios, these protections fail: permutation can be reversed via weight alignment~\cite{mm0.0}, and scaling can be undone by dividing with base model parameters (see Appendix~\ref{Attacking PaRaMS}).

\subsection{Threat Model}
In this paper, we consider two parties: the defender and the adversary. The defender fine-tunes a pretrained model on specialized data and releases it under a restrictive open-source license. The adversary, for their own benefit, acquires the open-source model and merges it with other models they control, all of which are fine-tuned from the same pretrained base. This scenario is reasonable since most fine-tuning workflows in deep learning begin with widely used pretrained models~\cite{han2024parameter,du2022survey}. The following are the details of the two parties. 

\highlight{Defender's Capability.} 
The defender can control their model and modify it to ensure protection. Specifically, the defender aims to design proactive defenses that preserve the model’s original performance while significantly degrading its performance when merged with other models.

\highlight{Adversary's Capability.} 
The adversary has limited data collection and computational resources, making it infeasible to train a high-quality target model on their own. Instead, they use model merging as a low-cost solution to acquire the defender’s specialized capabilities. The adversary has fine-tuned models based on the same pretrained model and can white-box access to all open-source resources, including the defender's open-source model, the pretrained base model, and defense strategy papers. They use these resources to optimize the performance of the merged model.

\section{Our Design: MergeBarrier}

This section introduces MergeBarrier, a proactive defense against model merging stealing.

\begin{tcolorbox}[colback=gray!20!white,colframe=black,left=2mm, right=2mm, top=1mm, bottom=1mm, boxsep=0pt]
\textbf{Key Idea:} MergeBarrier is based on the insight that successful model merging requires the expert models to reside within the same low-loss basin, thus forming a linear path where all intermediate points, acting as merged models, lie within the low-loss basin and retain high performance.
To prevent model merging, we transform the original model’s weights to displace it into a different loss basin, thereby eliminating the low-loss linear path, thus degrading the merged model's performance.

\end{tcolorbox}

\subsection{Overview}

\begin{figure}[t]
    \centering
    \includegraphics[width=1\linewidth]{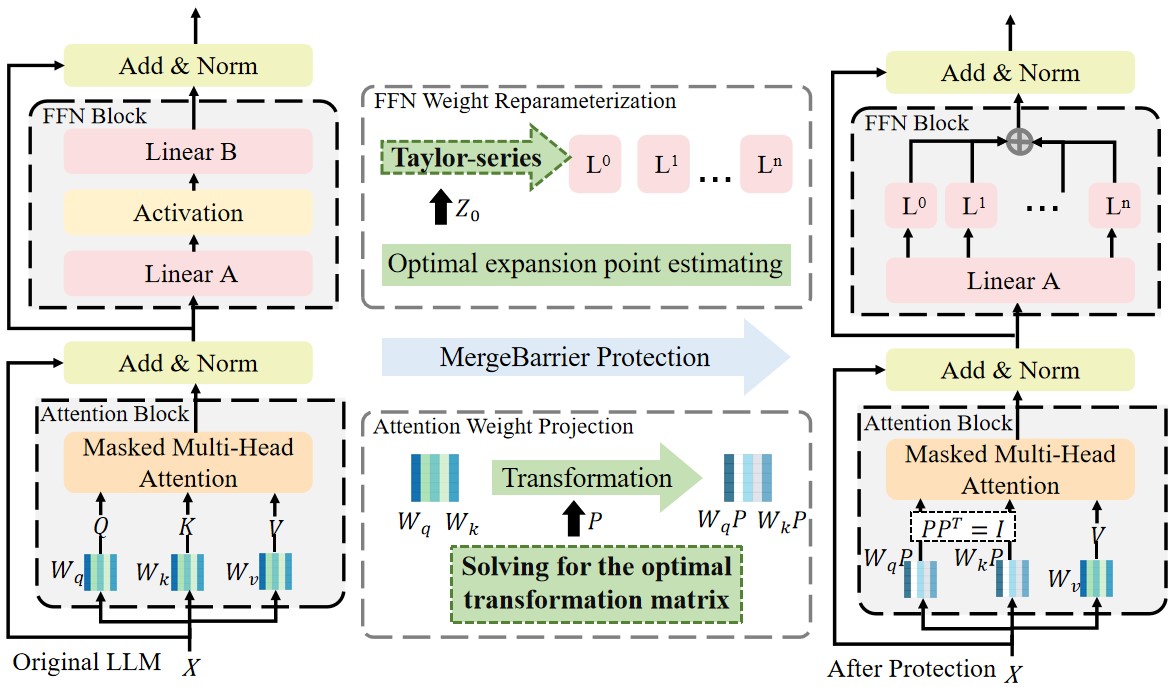}
    \caption{A pipeline of MergeBarrier. The method applies orthogonal projection to attention layers and reparameterization to FFN blocks to disrupt model merging, while preserving the model’s original performance.}
    \label{method}
\end{figure}

As shown in Figure~\ref{method}, MergeBarrier applies the key idea to the attention weight with projection and FFN weights with reparameterization, while preserving the model’s original performance.
Specifically, the \textit{attention weight projection} applies a shared orthogonal projection to both the query and key layers, thereby transforming the original parameter values and breaking the homogeneity with other homologous models. 
Furthermore, we optimize the orthogonal matrix to maximize the deviation between the merged model and the original low-loss region, thereby ensuring that the transformation is sufficiently effective.
After protection, due to the mathematical properties of the orthogonal matrix, its influence on the attention output is canceled out during computation, thereby preserving model performance.

For FFN blocks, MergeBarrier transforms the FFN weights by reparameterizing the feed-forward computation. We approximate the activation function using an activation function expansion. This transforms the original weights into a new set of functionally equivalent parameters, making the original weights inaccessible. As a result, attackers can no longer perform model merging via interpolation in the original weight space.
To preserve model performance, we select a local expansion point that minimizes the gap between the true input and its Taylor approximation. We further analyze the resulting remainder error and demonstrate that this noise acts as a built-in defense by transforming weight inversion into an NP-hard LWE problem~\cite{regev2009lattices}, making such attacks computationally infeasible.

\subsection{Attention Weight Projection}

\subsubsection{Orthogonal Transformation of Weight Matrix.}
We apply a shared orthogonal transformation to the query and key projection layer. Specifically, given an orthogonal projection matrix $P$, we apply it to the attention weights as follows:
\begin{equation}\label{ap}
{\footnotesize
\begin{aligned}
O &= \mathrm{softmax}\left(
    \frac{XW_q P P^{\top} W_k^{\top} X^{\top}}{\sqrt{c}}
    \right) XW_v \\
  &= \mathrm{softmax}\left(
    \frac{XW_q W_k^{\top} X^{\top}}{\sqrt{c}}
    \right) XW_v
\end{aligned}
}
\end{equation}

where $W_q$, $W_k$, and $W_v$ represent the weight matrices of the linear layers, and $X$ denotes the input sequence.
Since $P$ is an orthogonal matrix, it satisfies $PP^{\top} = I$, which ensures that the attention output remains unchanged for any such orthogonal transformation.

For attention variants such as Multi-Head Attention (MHA) and Group query Attention (GQA), we further prove in the appendix \ref{Proof of Theorem 1} that the above orthogonal transformation remains applicable, leading to the following theorem:
\begin{theorem}\label{th0}
    (a) In case of Multi-Head Attention, there exists a kind of orthogonal matrix $\hat{P}$ like $diag\{P_1, P_2, \cdots, P_n\}$, where $n$ represents the number of heads.
    
    (b) In case of Group Query Attention, there exists a kind of orthogonal matrix $\tilde{P}$ like $diag\{\hat{P}_1, \hat{P}_2, \cdots, \hat{P}_n\}$, where $n$ is equal to the number of query in a group.
\end{theorem}

\subsubsection{Optimizing the Orthogonal Matrix.}

To ensure that the merged model is pushed out of the original low-loss basin, we aim to find an orthogonal matrix $P$ that maximizes its distance from the basin.
Specifically, to represent the weight space, we jointly consider the query and key projection weights $W_q$ and $W_k$, as they together determine the attention output.
To locate the low-loss basin, we approximate it using the original model weights, i.e., $W_q W_k^\top$.
To express the merged model, we adopt weight averaging as a representative merging strategy. Therefore, the merged model after protection is represented as $\frac{1}{4}(W_{q}P+W_{q})(W_{k}P+W_{k})^{\top}$.
As a result, our objective can be formulated as maximizing the Frobenius norm of the difference between the merged model and the original basin:
\begin{equation}\label{object}
{\footnotesize
    \begin{aligned}
        \max_{P}\quad&\frac{1}{16}||(W_{q}P+W_{q})(W_{k}P+W_{k})^{\top}\\
        -&(W_{q}+W_{q})(W_{k}+W_{k})^{\top}||^2_{F}
    \end{aligned}
}
\end{equation}

The result is summarized in the following theorem, with the full proof provided in Appendix~\ref{Proof of Theorem 2}:
\begin{theorem}\label{th1}
(a) Maximizing Eq. (\ref{object}) is sufficient to maximize $||W_{q}(P-I)||^2_{F}+||W_{k}(P-I)||^2_{F}$.

(b) If $\lambda_i > 0$, then $(U^{\top}PU)_{ii} = -1$; if $\lambda_i < 0$, then $(U^{\top}PU)_{ii} = 1$.

(c) If $\lambda_i = 0$, then $(U^{\top}PU)_{ii}$ can take any value.
\end{theorem}
Ideally, the maximum perturbation occurs when $P = -I$, which flips all directions and maximizes the distance, but this would expose real model weights.
To solve this, we take a relaxed approach: set $(U^\top P U)_{ii} = -1$ for the top-$k$ eigen-directions (i.e., those with the largest $\Lambda$ values), and $1$ for the rest. By applying the inverse transformation, we obtain $P$ that satisfies the required conditions.

\subsubsection{Acceleration of Eigenvalue Decomposition.}

Directly computing the eigendecomposition of high-dimensional matrices like \(W_q^\top W_q\) is computationally expensive and impractical for large models, potentially making the method infeasible in practice. To address this, we employ Randomized Singular Value Decomposition (RSVD), which exploits the low-rank structure common in neural network weights. RSVD first projects the original matrix into a low-dimensional subspace using a random Gaussian matrix, then performs an SVD decomposition in this reduced-dimensional space, and finally reconstructs the approximation in the original space. This reduces the computational complexity from \(O(n^3)\) to \(O(n^2k + nk^2 + k^3)\), where \(k \ll n\), making RSVD well suited for large matrices. Details are provided in Appendix~\ref{Accelation}.

\subsection{FFN Weight Reparameterization}
Unlike attention blocks, FFN blocks lack a symmetric linear structure, making orthogonal transformations inapplicable. To address this, we adopt a reparameterization strategy tailored for FFN layers.
The key idea is to express the forward process using a new set of parameters, so the original weights are no longer explicitly present, thereby relocating the model in weight space and preventing model merging.
Specifically, we approximate the activation function using a polynomial expansion 
and replace the original FFN parameters with the corresponding polynomial coefficients.

\subsubsection{Reparameterizing FFN Weight by Polynomial expansion.}
For an FFN block, let $W$ and $c$ denote the weights and biases of the second linear layer, respectively. The output of this layer is denoted as $y$. We reparameterize it using a Taylor series as follows:
\begin{equation}\label{tayler}
{\scriptsize 
\begin{aligned}
    &y-c\\
    =&(y_1,\cdots,y_n)-(c_1,\cdots,c_n)\\
    =&(W_1\odot Act(z+b),W_2\odot Act(z+b),\cdots,W_n\odot Act(z+b))\\
    \approx&(W_1\odot \sum^{N}_{n=0}\frac{Act^{(n)}(z_0+b)}{n!},\cdots,W_n\odot \sum^{N}_{n=0}\frac{Act^{(n)}(z_0+b)}{n!})\\
    =&W\frac{Act^{(0)}(z_0+b)}{0!}(z_0-z)^0+\cdots+W\frac{Act^{(N)}(z_0+b)}{N!}(z_0-z)^N\\
    =&\hat{W}^0(z_0-z)^0+\cdots+\hat{W}^N(z_0-z)^N.
\end{aligned}
}
\end{equation}
Here, $W_i$ is i-th column of $W$. $z$ is the output of the first linear layer with bias $b$. The expansion point $z_0$ is set as the midpoint between $z_{max}$ and $z_{min}$ in the set $\mathcal{Z}$, which is constructed from the feature representations $z$ of training samples. The symbol $\odot$ denotes element-wise multiplication. 

Equation~\eqref{tayler} replaces the original weights with a polynomial approximation $\hat{W}^i$, enabling FFN computation without original weights.
Moreover, this approach is broadly applicable, supporting various activation functions commonly used in LLMs, such as $GELU(\cdot)$ \cite{vaswani2017attention} and $SiLU(\cdot)$ \cite{elfwing2018sigmoid}, as provided in Appendices~\ref{gelu} and~\ref{silu}. Additionally, the method is not limited to Taylor expansion—alternative bases, such as Hermite polynomials, can also be adopted.

\subsubsection{Discussion on Runtime Efficiency.}
\label{limitation}
MergeBarrier protects the MLP by modifying the model structure, which increases the computational cost of MLP blocks. However, in practice, the impact on runtime efficiency is minimal since computations at each order can run in parallel, allowing little to no increase in overall MLP block inference time.

\subsubsection{Analyzing the Taylor Remainder.}




The remainder term in a Taylor expansion inherently depends on higher-order derivatives and the expansion point, introducing noise that can be modeled probabilistically. Interestingly, our analysis reveals that this noise can act as a built-in safeguard against potential attacks.
For example, a potential attack strategy is to invert the Taylor expansion to recover the original weights. However, direct inversion is fundamentally infeasible: the system is underdetermined with more unknowns than equations, making its exact value inaccessible.
Even if the attacker fully leverages the base model parameters as prior knowledge during the inversion process, the attempt still fails (detailed in Appendix~\ref{LWE}). Specifically, the presence of the remainder term injects uncertainty into the relationship between the observed coefficients and the true weights. As a result, recovering the original weights reduces to solving a Learning With Errors (LWE) problem, which is widely regarded as NP-hard.

Additionally, we observe that the remainder-induced noise can serve as intrinsic noise that satisfies differential privacy (DP) (proof see Appendix \ref{Proof of DP}). Therefore, it enhances the weight's privacy, e.g., defending against membership inference attacks. 

\section{Experiments}

\subsection{Experimental Settings}
\label{Experimental Settings}
\subsubsection{Upstream LLMs.} 
We evaluate MergeBarrier by protecting expert models from different domains, including general language understanding, mathematical reasoning, and code generation.
We use LLaMA-2-13B~\cite{touvron2023llama} as the base model.
We evaluate three expert models, all fine-tuned from LLaMA-2-13B: WizardLM-13B~\cite{xu2023wizardlm} ($LM$) for general language understanding, WizardMath-13B~\cite{luo2023wizardmath} ($Math$) for mathematical reasoning, and LLaMA-2-13B-Code Alpaca~\cite{chaudhary2023code} ($Code$) for programming and code generation.

\subsubsection{Dataset and Metrics.} We evaluate the merged model on the representative dataset corresponding to the protected expert model's domain.
\textbf{(1) Instruction-following:} AlpacaEval\cite{li2023alpacaeval} tests a model’s ability to follow instructions and produce relevant responses. We report the win rate.
\textbf{(2) Math:} GSM8K\cite{cobbe2021training} contains grade-school math problems requiring multi-step reasoning and basic computation. Performance is measured by zero-shot accuracy.
\textbf{(3) Code:} MBPP~\cite{austin2021program} assesses Python code generation for programming tasks. We use pass@1, the percentage of problems solved correctly on the first attempt.

\subsubsection{Baselines.}
To compare our proposed methods, we selected two baselines, PaRaMS~\cite{junhao2025disrupting} and TaylorMLP~\cite{wang2024taylor}, as they represent the most pertinent approaches in the same threat model.
PaRaMS is specifically designed to counter model merging stealing, directly aligning with our goal. TaylorMLP aims to protect model weights while preserving utility, without relying on additional components.

\subsubsection{Evaluated Attack.}
Following the prior work~\cite{junhao2025disrupting}, we examine three representative model merging techniques as the basis for model merging stealing: Task Arithmetic~\cite{ilharco2022editing}, TIES-merging~\cite{yadav2023ties}, and DARE~\cite{yu2024language}. For defenses that modify model structures (e.g., MergeBarrier and TaylorMLP), attackers revert modified layers to the base model’s original weights before merging.
For PaRaMS, the attackers decode the protected model using the base model (Appendix~\ref{Attacking PaRaMS}) and apply the recovered weights. 


\begin{table*}[t]
\renewcommand{\arraystretch}{0.9}
\centering
\setlength{\tabcolsep}{3pt}  
\resizebox{2.1\columnwidth}{!}{
\begin{tabular}{ccccccccccccccccc}
\toprule
\multirow{2}{*}{\parbox{1cm}{Merging \\ methods}} & \multirow{2}{*}{\parbox{1cm}{Expert \\ models}} & \multicolumn{3}{c}{\textbf{Without protection}} & & \multicolumn{3}{c}{\textbf{MergeBarrier (Ours)}} & & \multicolumn{3}{c}{\textbf{PaRaMS}} & & \multicolumn{3}{c}{\textbf{TaylorMLP}} \\ 
\cline{3-5} \cline{7-9} \cline{11-13} \cline{15-17} \noalign{\smallskip} & & Alpaca & GSM8K & MBPP & & Alpaca & GSM8K & MBPP & &  Alpaca & GSM8K & MBPP & &  Alpaca & GSM8K & MBPP \\
\cline{1-17} \noalign{\smallskip}
\multirow{4}{*}{Task} 
& $LM$ \& $Math$ & 26.87\scriptsize\textcolor{gray!150}{$\pm$2.52} & 62.17\scriptsize\textcolor{gray!150}{$\pm$3.42} & - & & 0.14\scriptsize\textcolor{gray!150}{$\pm$0.07} & 0.00\scriptsize\textcolor{gray!150}{$\pm$0.00} & - & & 26.51\scriptsize\textcolor{gray!150}{$\pm$1.92} & 59.60\scriptsize\textcolor{gray!150}{$\pm$3.15} & - & & 22.73\scriptsize\textcolor{gray!150}{$\pm$0.64} & 55.25\scriptsize\textcolor{gray!150}{$\pm$2.31} & -\\
& $LM$ \& $Code$ & 40.85\scriptsize\textcolor{gray!150}{$\pm$2.41} & - & 35.60\scriptsize\textcolor{gray!150}{$\pm$1.87} & & 0.16\scriptsize\textcolor{gray!150}{$\pm$0.07} & - & 0.00\scriptsize\textcolor{gray!150}{$\pm$0.00} & & 40.80\scriptsize\textcolor{gray!150}{$\pm$2.23} & - & 35.00\scriptsize\textcolor{gray!150}{$\pm$1.56} & & 37.83\scriptsize\textcolor{gray!150}{$\pm$2.37} & - & 31.40\scriptsize\textcolor{gray!150}{$\pm$1.12} \\
& $Math$ \& $Code$ & - & 58.38\scriptsize\textcolor{gray!150}{$\pm$2.24} & 21.80\scriptsize\textcolor{gray!150}{$\pm$1.11} & & - & 0.00\scriptsize\textcolor{gray!150}{$\pm$0.00} & 0.00\scriptsize\textcolor{gray!150}{$\pm$0.00} & & - & 58.64\scriptsize\textcolor{gray!150}{$\pm$2.08} & 21.40\scriptsize\textcolor{gray!150}{$\pm$1.02} & & - & 49.64\scriptsize\textcolor{gray!150}{$\pm$2.87} & 19.00\scriptsize\textcolor{gray!150}{$\pm$0.93}\\
& \textit{Three Models} & 30.81\scriptsize\textcolor{gray!150}{$\pm$1.20} & 59.82\scriptsize\textcolor{gray!150}{$\pm$2.05} & 33.80\scriptsize\textcolor{gray!150}{$\pm$2.42} & & 3.11\scriptsize\textcolor{gray!150}{$\pm$0.91} & 5.11\scriptsize\textcolor{gray!150}{$\pm$0.27} & 0.00\scriptsize\textcolor{gray!150}{$\pm$0.00} & & 30.47\scriptsize\textcolor{gray!150}{$\pm$2.34} & 59.18\scriptsize\textcolor{gray!150}{$\pm$1.87} & 33.10\scriptsize\textcolor{gray!150}{$\pm$1.22} & & 27.83\scriptsize\textcolor{gray!150}{$\pm$1.05} & 53.11\scriptsize\textcolor{gray!150}{$\pm$3.64} & 30.80\scriptsize\textcolor{gray!150}{$\pm$1.08} \\
\cline{1-17} \noalign{\smallskip}
\multirow{4}{*}{TIES} 
& $LM$ \& $Math$ & 43.28\scriptsize\textcolor{gray!150}{$\pm$3.67} & 24.93\scriptsize\textcolor{gray!150}{$\pm$1.13} & - & & 3.85\scriptsize\textcolor{gray!150}{$\pm$1.08} & 3.75\scriptsize\textcolor{gray!150}{$\pm$0.97} & - & & 42.85\scriptsize\textcolor{gray!150}{$\pm$3.52} & 23.61\scriptsize\textcolor{gray!150}{$\pm$0.96} & - & & 35.07\scriptsize\textcolor{gray!150}{$\pm$2.24} & 21.94\scriptsize\textcolor{gray!150}{$\pm$0.91} & -\\
& $LM$ \& $Code$ & 44.62\scriptsize\textcolor{gray!150}{$\pm$3.83} & - & 18.00\scriptsize\textcolor{gray!150}{$\pm$0.95} & & 5.04\scriptsize\textcolor{gray!150}{$\pm$1.43} & - & 0.00\scriptsize\textcolor{gray!150}{$\pm$0.00} & & 45.28\scriptsize\textcolor{gray!150}{$\pm$1.95} & - & 7.50\scriptsize\textcolor{gray!150}{$\pm$0.34} & & 44.96\scriptsize\textcolor{gray!150}{$\pm$2.59} & - & 16.00\scriptsize\textcolor{gray!150}{$\pm$0.81} \\
& $Math$ \& $Code$ & - & 63.76\scriptsize\textcolor{gray!150}{$\pm$2.44} & 20.00\scriptsize\textcolor{gray!150}{$\pm$0.92} & & - & 6.94\scriptsize\textcolor{gray!150}{$\pm$0.37} & 0.00\scriptsize\textcolor{gray!150}{$\pm$0.00} & & - & 64.68\scriptsize\textcolor{gray!150}{$\pm$3.56} & 20.00\scriptsize\textcolor{gray!150}{$\pm$1.00} & & - & 45.09\scriptsize\textcolor{gray!150}{$\pm$2.82} & 16.80\scriptsize\textcolor{gray!150}{$\pm$0.74}\\
& \textit{Three Models} & 44.65\scriptsize\textcolor{gray!150}{$\pm$1.75} & 67.55\scriptsize\textcolor{gray!150}{$\pm$2.64} & 30.80\scriptsize\textcolor{gray!150}{$\pm$1.41} & & 6.20\scriptsize\textcolor{gray!150}{$\pm$1.54} & 9.81\scriptsize\textcolor{gray!150}{$\pm$2.21} & 0.00\scriptsize\textcolor{gray!150}{$\pm$0.00} & & 44.08\scriptsize\textcolor{gray!150}{$\pm$4.69} & 67.07\scriptsize\textcolor{gray!150}{$\pm$2.50} & 29.80\scriptsize\textcolor{gray!150}{$\pm$1.33} & & 42.87\scriptsize\textcolor{gray!150}{$\pm$1.61} & 51.94\scriptsize\textcolor{gray!150}{$\pm$2.02} & 24.80\scriptsize\textcolor{gray!150}{$\pm$1.15} \\
\cline{1-17} \noalign{\smallskip}
\multirow{4}{*}{\parbox{1cm}{DARE \\- Task}}
& $LM$ \& $Math$ & 33.42\scriptsize\textcolor{gray!150}{$\pm$1.47} & 60.17\scriptsize\textcolor{gray!150}{$\pm$2.36} & - & & 9.85\scriptsize\textcolor{gray!150}{$\pm$1.93} & 4.01\scriptsize\textcolor{gray!150}{$\pm$0.94} & - & & 32.75\scriptsize\textcolor{gray!150}{$\pm$2.51} & 61.70\scriptsize\textcolor{gray!150}{$\pm$2.41} & - & & 30.91\scriptsize\textcolor{gray!150}{$\pm$2.34} & 59.91\scriptsize\textcolor{gray!150}{$\pm$2.20} & -\\
& $LM$ \& $Code$ & 37.64\scriptsize\textcolor{gray!150}{$\pm$2.68} & - & 34.40\scriptsize\textcolor{gray!150}{$\pm$1.43} & & 9.71\scriptsize\textcolor{gray!150}{$\pm$0.84} & - & 0.00\scriptsize\textcolor{gray!150}{$\pm$0.00} & & 37.25\scriptsize\textcolor{gray!150}{$\pm$2.64} & - & 33.80\scriptsize\textcolor{gray!150}{$\pm$2.36} & & 34.88\scriptsize\textcolor{gray!150}{$\pm$3.33} & - & 26.80\scriptsize\textcolor{gray!150}{$\pm$1.10} \\
& $Math$ \& $Code$ & - & 59.37\scriptsize\textcolor{gray!150}{$\pm$2.22} & 23.80\scriptsize\textcolor{gray!150}{$\pm$1.15} & & - & 6.71\scriptsize\textcolor{gray!150}{$\pm$1.46} & 0.00\scriptsize\textcolor{gray!150}{$\pm$0.00} & & - & 58.43\scriptsize\textcolor{gray!150}{$\pm$2.04} & 23.20\scriptsize\textcolor{gray!150}{$\pm$0.47} & & - & 53.92\scriptsize\textcolor{gray!150}{$\pm$3.88} & 12.90\scriptsize\textcolor{gray!150}{$\pm$0.74}\\
& \textit{Three Models} & 35.83\scriptsize\textcolor{gray!150}{$\pm$2.41} & 51.82\scriptsize\textcolor{gray!150}{$\pm$1.92} & 34.60\scriptsize\textcolor{gray!150}{$\pm$2.32} & & 12.61\scriptsize\textcolor{gray!150}{$\pm$2.01} & 10.04\scriptsize\textcolor{gray!150}{$\pm$1.89} & 0.00\scriptsize\textcolor{gray!150}{$\pm$0.00} & & 29.99\scriptsize\textcolor{gray!150}{$\pm$2.24} & 60.07\scriptsize\textcolor{gray!150}{$\pm$2.18} & 34.60\scriptsize\textcolor{gray!150}{$\pm$3.40} & & 25.82\scriptsize\textcolor{gray!150}{$\pm$2.11} & 52.18\scriptsize\textcolor{gray!150}{$\pm$4.96} & 30.90\scriptsize\textcolor{gray!150}{$\pm$2.28} \\
\cline{1-17} \noalign{\smallskip}
\multirow{4}{*}{\parbox{1cm}{DARE \\- TIES}}
& $LM$ \& $Math$ & 45.84\scriptsize\textcolor{gray!150}{$\pm$2.74} & 30.39\scriptsize\textcolor{gray!150}{$\pm$1.32} & - & & 10.84\scriptsize\textcolor{gray!150}{$\pm$1.04} & 10.98\scriptsize\textcolor{gray!150}{$\pm$2.05} & - & & 46.27\scriptsize\textcolor{gray!150}{$\pm$2.69} & 29.30\scriptsize\textcolor{gray!150}{$\pm$1.28} & - & & 41.42\scriptsize\textcolor{gray!150}{$\pm$2.52} & 26.85\scriptsize\textcolor{gray!150}{$\pm$1.91} & -\\
& $LM$ \& $Code$ & 44.96\scriptsize\textcolor{gray!150}{$\pm$1.88} & - & 34.80\scriptsize\textcolor{gray!150}{$\pm$1.49} & & 10.58\scriptsize\textcolor{gray!150}{$\pm$2.00} & - & 0.80\scriptsize\textcolor{gray!150}{$\pm$0.21} & & 44.51\scriptsize\textcolor{gray!150}{$\pm$3.71} & - & 34.30\scriptsize\textcolor{gray!150}{$\pm$2.36} & & 34.95\scriptsize\textcolor{gray!150}{$\pm$2.41} & - & 35.40\scriptsize\textcolor{gray!150}{$\pm$2.48} \\
& $Math$ \& $Code$ & - & 58.78\scriptsize\textcolor{gray!150}{$\pm$2.15} & 17.80\scriptsize\textcolor{gray!150}{$\pm$0.81} & & - & 9.87\scriptsize\textcolor{gray!150}{$\pm$0.79} & 3.00\scriptsize\textcolor{gray!150}{$\pm$0.56} & & - & 58.18\scriptsize\textcolor{gray!150}{$\pm$2.04} & 16.80\scriptsize\textcolor{gray!150}{$\pm$0.78} & & - & 53.81\scriptsize\textcolor{gray!150}{$\pm$1.93} & 13.80\scriptsize\textcolor{gray!150}{$\pm$0.64}\\
& \textit{Three Models} & 45.71\scriptsize\textcolor{gray!150}{$\pm$3.77} & 49.05\scriptsize\textcolor{gray!150}{$\pm$1.94} & 25.60\scriptsize\textcolor{gray!150}{$\pm$1.03} & & 11.09\scriptsize\textcolor{gray!150}{$\pm$2.17} & 13.81\scriptsize\textcolor{gray!150}{$\pm$1.64} & 7.50\scriptsize\textcolor{gray!150}{$\pm$0.27} & & 46.42\scriptsize\textcolor{gray!150}{$\pm$2.85} & 48.41\scriptsize\textcolor{gray!150}{$\pm$3.82} & 25.70\scriptsize\textcolor{gray!150}{$\pm$2.01} & & 32.81\scriptsize\textcolor{gray!150}{$\pm$1.45} & 42.94\scriptsize\textcolor{gray!150}{$\pm$3.72} & 21.40\scriptsize\textcolor{gray!150}{$\pm$0.92} \\
\cline{1-17} \noalign{\smallskip}
\multicolumn{2}{c}{\textbf{Relative Mean Accuracy $\downarrow$}} &  & \textbf{8.27}$\times$ & & &  & \textbf{1.00}$\times$ &  & &   & \textbf{8.15}$\times$ &  & &   & \textbf{7.17}$\times$ & \\
\bottomrule
\end{tabular}
}
\caption{Security assessment of MergeBarrier in preventing model merging stealing. We report the accuracy (\%) of the merged model across different settings. 
Notably, in each setting, only one expert model is protected, and we exclusively report the merged model’s performance on its corresponding task.
The last row shows the average attack accuracy as a multiple of our method ($\times$).
}
\label{security}
\end{table*}

\subsection{Security}

\begin{figure}[t]
    \centering
    \includegraphics[width=0.95\linewidth]{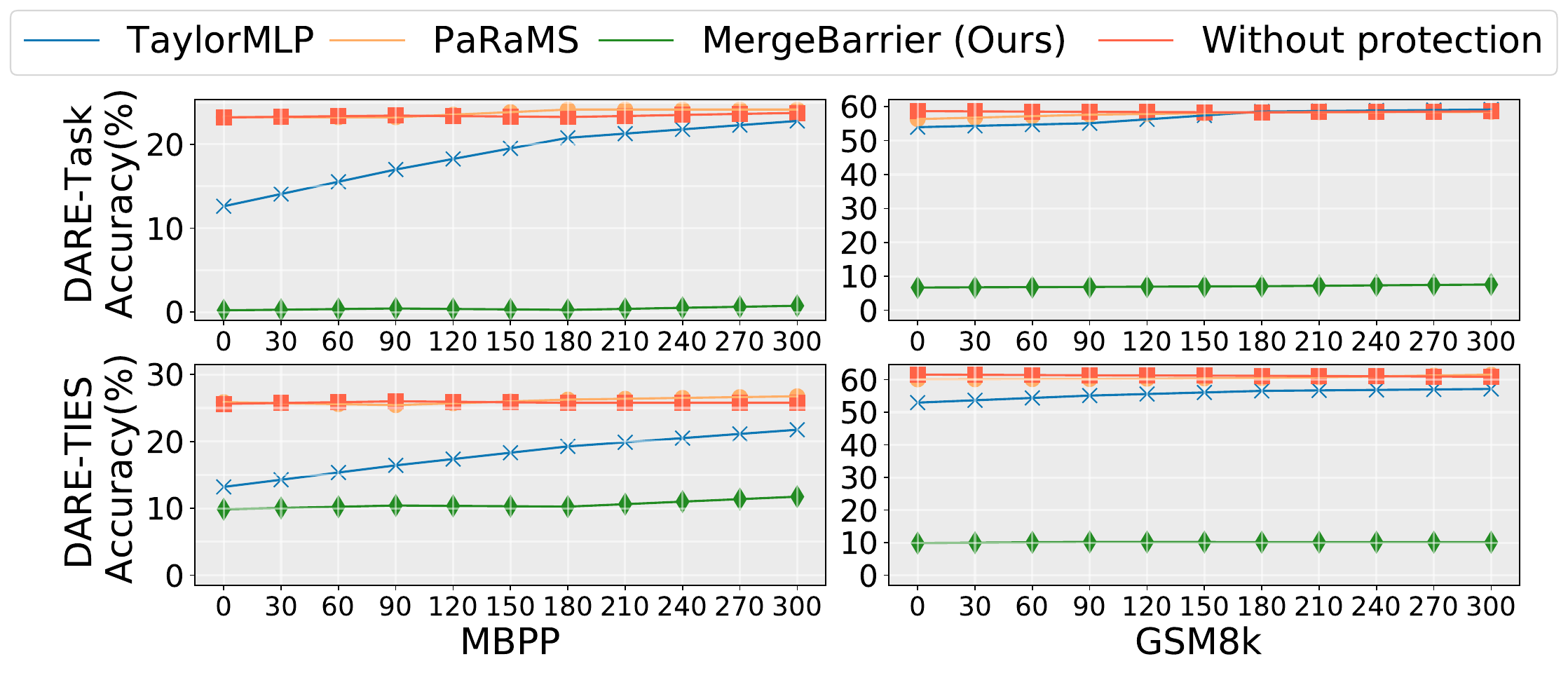}
    \caption{Defense effectiveness against model merging stealing with fine-tuning.}
    \label{experiments_4.4.2}
\end{figure}

\begin{figure}[t]
    \centering
    \includegraphics[width=0.95\linewidth]{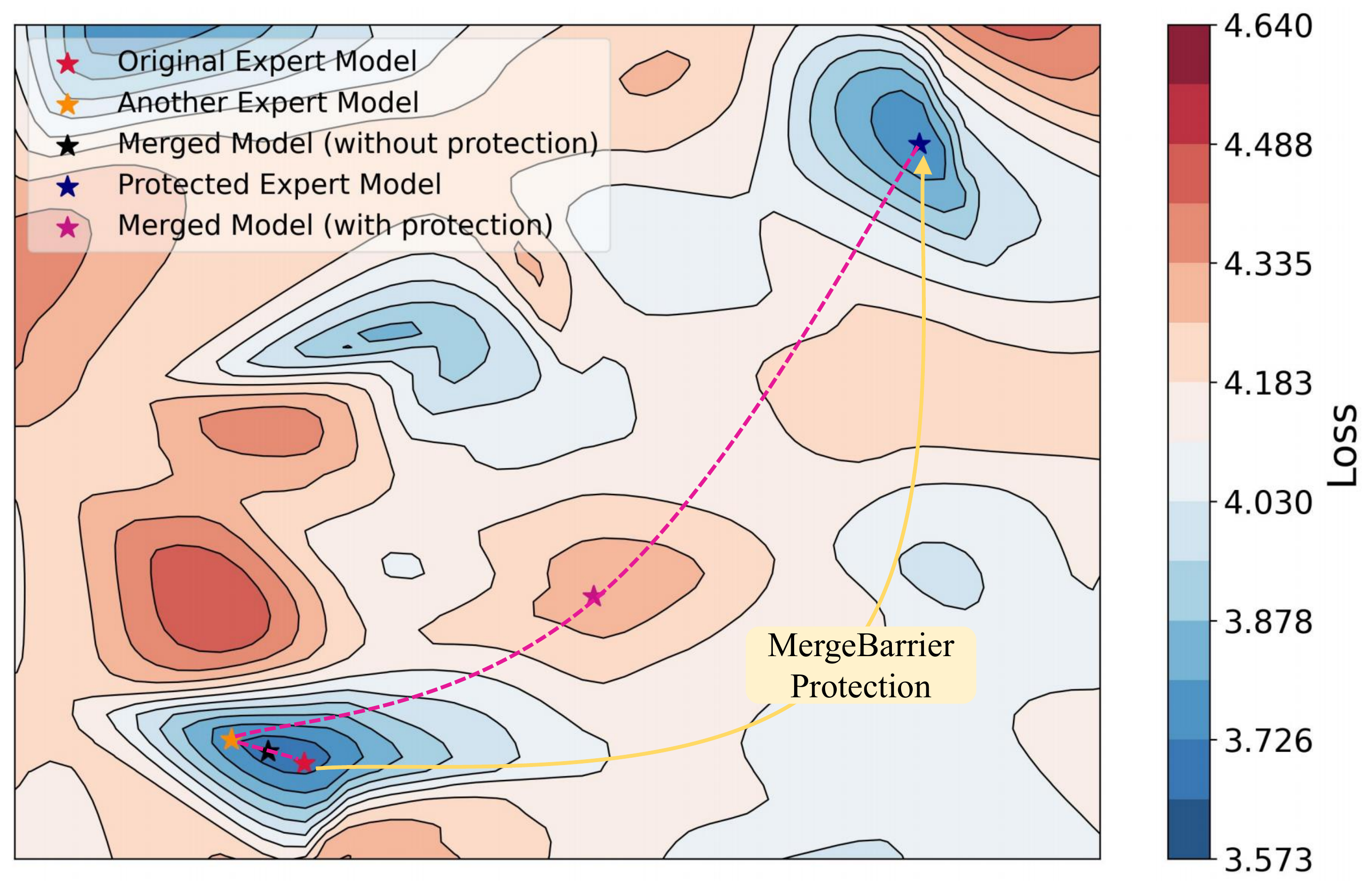}
    \caption{Loss landscape before and after our protection.}
    \label{LMC Analysis}
\end{figure}

\highlight{Security Against Model Merging Attacks.} 
In this subsection, we evaluate whether MergeBarrier effectively defends against model merging stealing.
To this end, we simulate an attacker who merges the protected model and reports the merged model’s performance on the corresponding task. If its task performance drops significantly, we consider the defense has successfully mitigated the attack.

As shown in Table~\ref{security}, MergeBarrier consistently suppresses the merged model’s performance across all merging settings. For example, in the \textit{Task} setting, MBPP accuracy drops to 0.00\% in both the “LM \& Code” and “Three Models” cases, and GSM8K accuracy in “LM \& Math” drops from 62.17\% to 0.00\%. Similar trends are observed under \textit{TIES} and \textit{DARE}.
To provide a holistic comparison, we compute the average merged model accuracy across all settings. MergeBarrier achieves the lowest score (1.00×), indicating near-complete suppression of merging-based theft. In contrast, PaRaMS and TaylorMLP fail to provide effective protection, with the merged models retaining up to 31.10\% and 30.80\% MBPP accuracy, and significantly higher average scores of 8.15× and 7.17×, respectively.

This strong protection stems from MergeBarrier’s comprehensive protection. Specifically, attention projection disrupts cross-model weight alignment, while FFN reparameterization conceals true weight.
In contrast, PaRaMS and TaylorMLP either fail under adaptive attacks or safeguard only a limited subset of weights.

\subsubsection*{Security Against Stealing with Fine-Tuning.}

In this section, we consider a stronger attacker who further fine-tunes the merged model with limited labeled data, increasing its task performance, thus increasing the threat. We report the task performance of different defenses under this enhanced model merging stealing scenario, across two merging strategies and two target tasks.
As shown in Figure~\ref{experiments_4.4.2}, MergeBarrier remains highly robust: even with more labeled data, the merged model’s performance improves only slightly. In contrast, TaylorMLP is easily compromised: minor fine-tuning significantly improves task performance. This strong resistance stems from MergeBarrier’s comprehensive protection over both attention and FFN parameters.


\begin{table*}[t]
\centering
\renewcommand{\arraystretch}{0.9}
\resizebox{0.66\textwidth}{!}{
\begin{tabular}{cccccccc}
\toprule
\multirow{2}{*}{\textbf{Models}} & \multicolumn{3}{c}{\textbf{Accuracy $\uparrow$}} & & \multicolumn{3}{c}{\textbf{Sharpness $\downarrow$}} \\ 
\cline{2-4} \cline{6-8} \noalign{\smallskip} & Alpaca & GSM8K & MBPP & & Alpaca & GSM8K & MBPP \\
\midrule
 WizardLM-13B & 45.59 / 45.53 & -  & - &  & 0.53 / 0.54 & -  & - \\
 WizardMath-13B & - & 58.76 / 57.88  & - &  & - & 0.57 / 0.58  & - \\
LLaMA-2-13B-Code  & - & -  & 27.80 / 27.80 &  & - & -  & 0.63 / 0.65 \\
\bottomrule
\end{tabular}
}
\caption{
The performance comparison between the original model ($M$) and the MergeBarrier-protected model ($M'$). The result is presented in the form of $M$ / $M'$.
}
\label{Accuracy Loss}
\end{table*}

\begin{table*}[t]
\renewcommand{\arraystretch}{0.9}
\centering
\setlength{\tabcolsep}{3pt}  
\resizebox{2.1\columnwidth}{!}{
\begin{tabular}{ccccccccccccccccc}
\toprule
\multirow{2}{*}{\parbox{1cm}{Merging \\ methods}} & \multirow{2}{*}{\parbox{1cm}{Expert \\ models}} & \multicolumn{3}{c}{\textbf{Without protection}} & & \multicolumn{3}{c}{\textbf{MergeBarrier (Ours)}} & & \multicolumn{3}{c}{\textbf{MergeBarrier$_{Att}$}} & & \multicolumn{3}{c}{\textbf{MergeBarrier$_{FFN}$}} \\ 
\cline{3-5} \cline{7-9} \cline{11-13} \cline{15-17} \noalign{\smallskip} & & Alpaca & GSM8K & MBPP & & Alpaca & GSM8K & MBPP & &  Alpaca & GSM8K & MBPP & &  Alpaca & GSM8K & MBPP \\
\cline{1-17} \noalign{\smallskip}
\multirow{4}{*}{Task} 
& $LM$ \& $Math$ & 26.87\scriptsize\textcolor{gray!150}{$\pm$2.52} & 62.17\scriptsize\textcolor{gray!150}{$\pm$3.42} & - & & 0.14\scriptsize\textcolor{gray!150}{$\pm$0.07} & 0.00\scriptsize\textcolor{gray!150}{$\pm$0.00} & - & & 12.45\scriptsize\textcolor{gray!150}{$\pm$1.24} & 21.20\scriptsize\textcolor{gray!150}{$\pm$1.06} & - & & 23.61\scriptsize\textcolor{gray!150}{$\pm$1.42} & 56.12\scriptsize\textcolor{gray!150}{$\pm$2.83} & -\\
& $LM$ \& $Code$ & 40.85\scriptsize\textcolor{gray!150}{$\pm$2.41} & - & 35.60\scriptsize\textcolor{gray!150}{$\pm$1.87} & & 0.16\scriptsize\textcolor{gray!150}{$\pm$0.07} & - & 0.00\scriptsize\textcolor{gray!150}{$\pm$0.00} & & 14.61\scriptsize\textcolor{gray!150}{$\pm$1.34} & - & 11.90\scriptsize\textcolor{gray!150}{$\pm$0.87} & & 35.66\scriptsize\textcolor{gray!150}{$\pm$1.78} & - & 30.90\scriptsize\textcolor{gray!150}{$\pm$1.52} \\
& $Math$ \& $Code$ & - & 58.38\scriptsize\textcolor{gray!150}{$\pm$2.24} & 21.80\scriptsize\textcolor{gray!150}{$\pm$1.11} & & - & 0.00\scriptsize\textcolor{gray!150}{$\pm$0.00} & 0.00\scriptsize\textcolor{gray!150}{$\pm$0.00} & & - & 13.26\scriptsize\textcolor{gray!150}{$\pm$0.92} & 15.30\scriptsize\textcolor{gray!150}{$\pm$0.78} & & - & 49.83\scriptsize\textcolor{gray!150}{$\pm$2.43} & 18.80\scriptsize\textcolor{gray!150}{$\pm$1.11}\\
& \textit{Three Models} & 30.81\scriptsize\textcolor{gray!150}{$\pm$1.20} & 59.82\scriptsize\textcolor{gray!150}{$\pm$2.05} & 33.80\scriptsize\textcolor{gray!150}{$\pm$2.42} & & 3.11\scriptsize\textcolor{gray!150}{$\pm$0.91} & 5.11\scriptsize\textcolor{gray!150}{$\pm$0.27} & 0.00\scriptsize\textcolor{gray!150}{$\pm$0.00} & & 23.26\scriptsize\textcolor{gray!150}{$\pm$1.56} & 25.35\scriptsize\textcolor{gray!150}{$\pm$1.42} & 24.10\scriptsize\textcolor{gray!150}{$\pm$1.23} & & 28.62\scriptsize\textcolor{gray!150}{$\pm$1.64} & 42.56\scriptsize\textcolor{gray!150}{$\pm$2.16} & 31.60\scriptsize\textcolor{gray!150}{$\pm$1.52} \\
\cline{1-17} \noalign{\smallskip}
\multirow{4}{*}{TIES} 
& $LM$ \& $Math$ & 43.28\scriptsize\textcolor{gray!150}{$\pm$3.67} & 24.93\scriptsize\textcolor{gray!150}{$\pm$1.13} & - & & 3.85\scriptsize\textcolor{gray!150}{$\pm$1.08} & 3.75\scriptsize\textcolor{gray!150}{$\pm$0.97} & - & & 15.12\scriptsize\textcolor{gray!150}{$\pm$1.12} & 18.93\scriptsize\textcolor{gray!150}{$\pm$1.41} & - & & 34.49\scriptsize\textcolor{gray!150}{$\pm$2.47} & 22.62\scriptsize\textcolor{gray!150}{$\pm$1.33} & -\\
& $LM$ \& $Code$ & 44.62\scriptsize\textcolor{gray!150}{$\pm$3.83} & - & 18.00\scriptsize\textcolor{gray!150}{$\pm$0.95} & & 5.04\scriptsize\textcolor{gray!150}{$\pm$1.43} & - & 0.00\scriptsize\textcolor{gray!150}{$\pm$0.00} & & 11.21\scriptsize\textcolor{gray!150}{$\pm$0.84} & - & 6.30\scriptsize\textcolor{gray!150}{$\pm$0.49} & & 44.70\scriptsize\textcolor{gray!150}{$\pm$2.12} & - & 16.60\scriptsize\textcolor{gray!150}{$\pm$0.91} \\
& $Math$ \& $Code$ & - & 63.76\scriptsize\textcolor{gray!150}{$\pm$2.44} & 20.00\scriptsize\textcolor{gray!150}{$\pm$0.92} & & - & 6.94\scriptsize\textcolor{gray!150}{$\pm$0.37} & 0.00\scriptsize\textcolor{gray!150}{$\pm$0.00} & & - & 44.22\scriptsize\textcolor{gray!150}{$\pm$1.83} & 15.30\scriptsize\textcolor{gray!150}{$\pm$0.72} & & - & 61.33\scriptsize\textcolor{gray!150}{$\pm$2.65} & 17.00\scriptsize\textcolor{gray!150}{$\pm$0.89}\\
& \textit{Three Models} & 44.65\scriptsize\textcolor{gray!150}{$\pm$1.75} & 67.55\scriptsize\textcolor{gray!150}{$\pm$2.64} & 30.80\scriptsize\textcolor{gray!150}{$\pm$1.41} & & 6.20\scriptsize\textcolor{gray!150}{$\pm$1.54} & 9.81\scriptsize\textcolor{gray!150}{$\pm$2.21} & 0.00\scriptsize\textcolor{gray!150}{$\pm$0.00} & & 22.93\scriptsize\textcolor{gray!150}{$\pm$1.52} & 49.18\scriptsize\textcolor{gray!150}{$\pm$2.47} & 17.50\scriptsize\textcolor{gray!150}{$\pm$0.83} & & 42.72\scriptsize\textcolor{gray!150}{$\pm$2.12} & 52.12\scriptsize\textcolor{gray!150}{$\pm$2.39} & 25.40\scriptsize\textcolor{gray!150}{$\pm$1.21} \\
\cline{1-17} \noalign{\smallskip}
\multirow{4}{*}{\parbox{1cm}{DARE \\- Task}}
& $LM$ \& $Math$ & 33.42\scriptsize\textcolor{gray!150}{$\pm$1.47} & 60.17\scriptsize\textcolor{gray!150}{$\pm$2.36} & - & & 9.85\scriptsize\textcolor{gray!150}{$\pm$1.93} & 4.01\scriptsize\textcolor{gray!150}{$\pm$0.94} & - & & 12.71\scriptsize\textcolor{gray!150}{$\pm$1.08} & 11.51\scriptsize\textcolor{gray!150}{$\pm$0.92} & - & & 31.37\scriptsize\textcolor{gray!150}{$\pm$1.77} & 59.87\scriptsize\textcolor{gray!150}{$\pm$2.48} & -\\
& $LM$ \& $Code$ & 37.64\scriptsize\textcolor{gray!150}{$\pm$2.68} & - & 34.40\scriptsize\textcolor{gray!150}{$\pm$1.43} & & 9.71\scriptsize\textcolor{gray!150}{$\pm$0.84} & - & 0.00\scriptsize\textcolor{gray!150}{$\pm$0.00} & & 10.23\scriptsize\textcolor{gray!150}{$\pm$0.74} & - & 2.80\scriptsize\textcolor{gray!150}{$\pm$0.21} & & 35.14\scriptsize\textcolor{gray!150}{$\pm$1.56} & - & 26.60\scriptsize\textcolor{gray!150}{$\pm$1.33} \\
& $Math$ \& $Code$ & - & 59.37\scriptsize\textcolor{gray!150}{$\pm$2.22} & 23.80\scriptsize\textcolor{gray!150}{$\pm$1.15} & & - & 6.71\scriptsize\textcolor{gray!150}{$\pm$1.46} & 0.00\scriptsize\textcolor{gray!150}{$\pm$0.00} & & - & 11.00\scriptsize\textcolor{gray!150}{$\pm$0.93} & 6.30\scriptsize\textcolor{gray!150}{$\pm$0.47} & & - & 53.90\scriptsize\textcolor{gray!150}{$\pm$2.31} & 12.50\scriptsize\textcolor{gray!150}{$\pm$0.69}\\
& \textit{Three Models} & 35.83\scriptsize\textcolor{gray!150}{$\pm$2.41} & 51.82\scriptsize\textcolor{gray!150}{$\pm$1.92} & 34.60\scriptsize\textcolor{gray!150}{$\pm$2.32} & & 12.61\scriptsize\textcolor{gray!150}{$\pm$2.01} & 10.04\scriptsize\textcolor{gray!150}{$\pm$1.89} & 0.00\scriptsize\textcolor{gray!150}{$\pm$0.00} & & 26.92\scriptsize\textcolor{gray!150}{$\pm$1.46} & 20.56\scriptsize\textcolor{gray!150}{$\pm$1.24} & 9.70\scriptsize\textcolor{gray!150}{$\pm$0.65} & & 25.96\scriptsize\textcolor{gray!150}{$\pm$1.68} & 52.26\scriptsize\textcolor{gray!150}{$\pm$2.36} & 30.50\scriptsize\textcolor{gray!150}{$\pm$1.41} \\
\cline{1-17} \noalign{\smallskip}
\multirow{4}{*}{\parbox{1cm}{DARE \\- TIES}}
& $LM$ \& $Math$ & 45.84\scriptsize\textcolor{gray!150}{$\pm$2.74} & 30.39\scriptsize\textcolor{gray!150}{$\pm$1.32} & - & & 10.84\scriptsize\textcolor{gray!150}{$\pm$1.04} & 10.98\scriptsize\textcolor{gray!150}{$\pm$2.05} & - & & 24.12\scriptsize\textcolor{gray!150}{$\pm$1.87} & 23.24\scriptsize\textcolor{gray!150}{$\pm$1.68} & - & & 41.90\scriptsize\textcolor{gray!150}{$\pm$2.28} & 27.44\scriptsize\textcolor{gray!150}{$\pm$1.42} & -\\
& $LM$ \& $Code$ & 44.96\scriptsize\textcolor{gray!150}{$\pm$1.88} & - & 34.80\scriptsize\textcolor{gray!150}{$\pm$1.49} & & 10.58\scriptsize\textcolor{gray!150}{$\pm$2.00} & - & 0.80\scriptsize\textcolor{gray!150}{$\pm$0.21} & & 24.67\scriptsize\textcolor{gray!150}{$\pm$1.74} & - & 12.30\scriptsize\textcolor{gray!150}{$\pm$0.83} & & 34.76\scriptsize\textcolor{gray!150}{$\pm$1.91} & - & 29.20\scriptsize\textcolor{gray!150}{$\pm$1.37} \\
& $Math$ \& $Code$ & - & 58.78\scriptsize\textcolor{gray!150}{$\pm$2.15} & 17.80\scriptsize\textcolor{gray!150}{$\pm$0.81} & & - & 9.87\scriptsize\textcolor{gray!150}{$\pm$0.79} & 3.00\scriptsize\textcolor{gray!150}{$\pm$0.56} & & - & 20.14\scriptsize\textcolor{gray!150}{$\pm$1.41} & 9.40\scriptsize\textcolor{gray!150}{$\pm$0.62} & & - & 53.52\scriptsize\textcolor{gray!150}{$\pm$2.14} & 15.10\scriptsize\textcolor{gray!150}{$\pm$0.85}\\
& \textit{Three Models} & 45.71\scriptsize\textcolor{gray!150}{$\pm$3.77} & 49.05\scriptsize\textcolor{gray!150}{$\pm$1.94} & 25.60\scriptsize\textcolor{gray!150}{$\pm$1.03} & & 11.09\scriptsize\textcolor{gray!150}{$\pm$2.17} & 13.81\scriptsize\textcolor{gray!150}{$\pm$1.64} & 7.50\scriptsize\textcolor{gray!150}{$\pm$0.27} & & 28.33\scriptsize\textcolor{gray!150}{$\pm$1.58} & 35.93\scriptsize\textcolor{gray!150}{$\pm$1.87} & 12.70\scriptsize\textcolor{gray!150}{$\pm$0.78} & & 33.20\scriptsize\textcolor{gray!150}{$\pm$1.91} & 43.04\scriptsize\textcolor{gray!150}{$\pm$2.18} & 20.90\scriptsize\textcolor{gray!150}{$\pm$1.09} \\
\cline{1-17} \noalign{\smallskip}
\multicolumn{2}{c}{\textbf{Relative Mean Accuracy $\downarrow$}} &  & \textbf{8.27}$\times$ & & &  & \textbf{1.00}$\times$ &  & &   & \textbf{3.79}$\times$ &  & &   & \textbf{7.19}$\times$ & \\
\bottomrule
\end{tabular}
}
\caption{Security assessment of MergeBarrier and its variants in preventing model merging stealing.
}
\label{Ablation study}
\end{table*}

\subsection{LMC Analysis}

We analyze how MergeBarrier affects Linear Mode Connectivity (LMC) by constructing weight-loss plots to visualize the relationship between models in weight space. Specifically, we apply PCA to reduce the dimensionality of model weights and project them into a 2D space, where we report the original loss values corresponding to each model in weight space.
As shown in Figure \ref{LMC Analysis}, before protection, the original model and the another expert lie within the same low-loss basin, so the linear path connecting them also remains in this basin. This ensures that the merged model, which is located along the linear path, also resides in a low-loss region and thus maintains good performance.
After MergeBarrier protection, the original model is relocated to a different basin, with no low-loss path connecting it to the another expert model. 
In other words, LMC has been disrupted, and any merged model formed along this path will suffer from high loss and degraded performance.


\subsection{Model Performance}

We assess MergeBarrier’s impact by comparing the performance of the protected model with the original model using both \textbf{accuracy} and \textbf{sharpness} metrics. Accuracy reflects performance under ideal conditions, where the model is evaluated with its exact trained weights. However, real-world deployments often introduce slight weight deviations, e.g., due to quantization or low-precision hardware, that may degrade performance. As MergeBarrier modifies model weights, it may affect the model's robustness to such deviations. To account for this, we also measure sharpness (detailed in Appendix~\ref{sharpness}), which quantifies the loss sensitivity to small parameter deviations. A lower sharpness indicates flatter minima and greater weight robustness.

As shown in Table~\ref{Accuracy Loss}, MergeBarrier introduces a negligible impact on both metrics. Accuracy is preserved (e.g., LLaMA-2-13B-Code: 27.80\% vs. 27.80\%), and sharpness remains stable (e.g., GSM8K: 0.57 to 0.58). 
These strong results stem from our targeted design: orthogonal projection in attention and functionally equivalent reparameterization in FFN are both crafted to preserve performance.


\subsection{Ablation Study}
We assess the individual contribution of each MergeBarrier component by comparing the full method with two variants: MergeBarrier${Att}$ (only attention projection) and MergeBarrier${FFN}$ (only FFN reparameterization). 
As shown in Table~\ref{Ablation study}, individually protecting either the attention or FFN module provides only limited defense against model merging attacks. Specifically, the full MergeBarrier achieves the lowest relative mean accuracy for the merged model (1.00×), while the attention-only and FFN-only variants reach 3.79× and 7.19×, respectively. 
These results demonstrate that robust protection against model merging requires the joint deployment of both attention projection and FFN reparameterization. In contrast, partial protection leaves significant vulnerabilities exploitable by attackers. This ablation study thus empirically validates the necessity and effectiveness of holistic defense in MergeBarrier.


\section{Conclusion}
In this work, we address the emerging threat of model merging stealing, where unauthorized users exploit proprietary models through merging. We highlight key limitations in current defense strategies and stress the need for secure, compatible methods that do not compromise model performance. To this end, we introduce MergeBarrier, a proactive defense mechanism that prevents unauthorized merging without impacting model utility.
Extensive experiments show that MergeBarrier outperforms existing methods, providing strong protection with minimal accuracy loss. Theoretical analysis also confirms that these transformations ensure differential privacy.
In conclusion, MergeBarrier is an effective solution that provides model owners with the means to safeguard their proprietary LLMs.

\newpage

\section*{Acknowledgements}
This work was supported by the Key Project of the National Natural Science Foundation of China under Grant no. 62536007, the Zhejiang Province Science Foundation under Grant no. LD24F020002 and the Zhejiang Province's 2025 “Leading Goose + X" Science and Technology Plan under Grant no. 2025C02034.

\bibliography{aaai2026}
\newpage
\appendix

\clearpage

\section{Appendix / Model Merging Techniques}\label{model merging techniques}

\highlight{Task Arithmetic.} Ilharco et al.~\cite{ilharco2022editing} introduce a merging approach that captures the difference between expert and base models using delta weights. This method constructs a merged model by computing a set of transformation vectors that encapsulate variations in performance. By applying a weighted combination of these vectors to the base model, the approach integrates knowledge from multiple sources while maintaining adaptability to different tasks.

\highlight{TIES-merging.} Yadav et al.~\cite{yadav2023ties} address challenges in model merging caused by conflicting parameter values across different models, which often degrade overall performance. Their approach consists of three sequential steps: refining the task vectors by retaining only the most influential components, determining the dominant sign for each parameter through aggregation, and forming the final merged representation by aligning values with the chosen signs. This structured methodology minimizes redundancy and enhances consistency in the merged model.

\highlight{DARE.} Drop And REscale (DARE)~\cite{yu2024language} is a pre-processing technique aimed at improving merging performance by selectively pruning weights and rescaling the remaining ones. It enhances model sparsity by probabilistically retaining a subset of weights while nullifying others, thereby reducing conflicts between expert models. To counterbalance the dropped values, the preserved weights are rescaled, ensuring stable performance. This technique operates independently of specific merging algorithms, making it compatible with other vector-based fusion methods for flexible integration.

\section{Appendix / Supplemental Material}


\subsection{Proof of Theorem \ref{th0}}
\label{Proof of Theorem 1}

For Multi-Head Attention case, we compute attention as follow:

\begin{equation}
    \begin{aligned}
        Q_i&=XW_{q_i},K=XW_{k_i}\\
        Attention&=\sum_{i=1}^n softmax(\frac{XW_qW_iW_i^{\top}W_k^{\top}X^{\top}}{\sqrt{c}}),
    \end{aligned}
\end{equation}

where $W_i$ refers to the slicing operation of the i-th head, that is, taking the weight $W_q$ row or column corresponding to this head. After parameter transformation, we have

\begin{equation}
    \begin{aligned}
        Attention&=\sum_{i=1}^n softmax(\frac{XW_qPW_iW_i^{\top}P^{\top}W_k^{\top}X^{\top}}{\sqrt{c}}).
    \end{aligned}
\end{equation}

Therefore, we need to prove

\begin{equation}
W_qPW_iW_i^{\top}P^{\top}W_k^{\top}=W_qW_iW_i^{\top}W_k^{\top}.
\end{equation}

We consider the attention of two heads, then for first head, we have

\begin{equation}\label{qiwang}
    \begin{aligned}
    &\begin{pmatrix}
        P_1 & 0 \\ 0 & P_2
    \end{pmatrix}
    \begin{pmatrix}
        I & 0 \\ 0 & 0
    \end{pmatrix}
    \begin{pmatrix}
        I & 0 \\ 0 & 0
    \end{pmatrix}
    \begin{pmatrix}
        P_1^{\top} & 0 \\ 0 & P_2^{\top}
    \end{pmatrix}
    \\=&
    \begin{pmatrix}
        P_1 & 0 \\ 0 & 0
    \end{pmatrix}
    \begin{pmatrix}
        P_1^{\top} & 0 \\ 0 & 0
    \end{pmatrix}.
    \end{aligned}
\end{equation}

Therefore, if $P_1$ is orthogonal matrix, we will have Eq. (\ref{qiwang}) established. The proof for the other heads is similar, so we have Theorem $\ref{th0}$ (a) holds.

For Group Query Attention case, we compute attention as follow:

\begin{equation}
Attention=softmax(\frac{XW_q\begin{pmatrix}I,I\end{pmatrix}^{\top}W_k^{\top}X^{\top}}{\sqrt{c}}).
\end{equation}

For the simplicity of the proof, we let the key in a group be replicated twice. $\begin{pmatrix}I,I\end{pmatrix}^{\top}$ means to copy the key in this query group, that is, to double the row dimension of $W_k$. After parameter transformation, we have

\begin{equation}
Attention=softmax(\frac{XW_q\tilde{P}\begin{pmatrix}I,I\end{pmatrix}^{\top}\tilde{P}^{\top}W_k^{\top}X^{\top}}{\sqrt{c}}).
\end{equation}

It is worth noting that $\tilde{P}$ is $diag\{\hat{P_1},\hat{P_2}\}$, because the dimensions of $Q$ and $K$ are different. Therefore, we need to prove

\begin{equation}\label{qi} W_q\tilde{P}\begin{pmatrix}I,I\end{pmatrix}^{\top}\tilde{P}^{\top}W_k^{\top}=W_q\begin{pmatrix}I,I\end{pmatrix}^{\top}W_k^{\top}.
\end{equation}

Starting from the left of Eq. \ref{qi}, we have

\begin{equation}
    \begin{aligned}
        &\begin{pmatrix}
            W_{q_1} & W_{q_2}\\
            W_{q_3} & W_{q_4}
        \end{pmatrix}\begin{pmatrix}
            \hat{P_1} & 0\\
            0 & \hat{P_2}
        \end{pmatrix}
        \begin{pmatrix}
            I\\ I
        \end{pmatrix}\begin{pmatrix}
            \hat{P_1}^{\top} & 0\\
            0 & \hat{P_2}^{\top}
        \end{pmatrix}
        W_k\\
        =&\begin{pmatrix}
            (W_{q_1}+W_{q_2})\hat{P_1}\hat{P_1}^{\top}W_k\\
            (W_{q_3}+W_{q_4})\hat{P_2}\hat{P_2}^{\top}W_k
        \end{pmatrix}\\
        =&\begin{pmatrix}
            (W_{q_1}+W_{q_2})W_k\\
            (W_{q_3}+W_{q_4})W_k
        \end{pmatrix}\\
        =&W_q\begin{pmatrix}I,I\end{pmatrix}^{\top}W_k^{\top}.
    \end{aligned}
\end{equation}

Therefore, if $\hat{P_1}$ and $\hat{P_2}$ are both orthogonal matrix, then Eq. \ref{qi} holds. For the multi-head case in GQA, the proof is similar and can be easily extended from the single-head case.

\subsection{Proof of Theorem \ref{th1}}
\label{Proof of Theorem 2}

Our objective is
\begin{equation}
    \begin{aligned}
        \max_{P}\quad&\frac{1}{16}\left\|(W_{q_1}P+W_{q_2})(W_{k_1}P+W_{k_2})^{\top}\right.\\
        -&\left.(W_{q_1}+W_{q_2})(W_{k_1}+W_{k_2})^{\top}\right\|^2_{F} \\
        =&\left\|W_{q_1}PW_{k_2}^{\top}+W_{q_2}P^{\top}W_{k_1}^{\top}-W_{q_1}W_{k_2}^{\top}-W_{q_2}W_{k_1}^{\top}\right\|^2_{F} \\
        =&\left\|W_{q_1}(P-I)W_{k_2}^{\top}+W_{q_2}(P-I)^{\top}W_{k_1}^{\top}\right\|^2_{F}.
    \end{aligned}
\end{equation}

In this formulation, $W_{q_1}$ and $W_{k_1}$ represent the protected model’s weights, which we aim to secure against merging. In contrast, $W_{q_2}$ and $W_{k_2}$ correspond to an arbitrary set of weights from an external model involved in the merging process. The two resulting terms in the loss function, $W_{q_1}(P-I)W_{k_2}^{\top}$ and $W_{q_2}(P-I)^{\top}W_{k_1}^{\top}$, exhibit structurally similar forms. 

However, a key observation is that $W_{q_2}$ and $W_{k_2}$ are arbitrary and unconstrained in practice, and can be viewed as independent of the protected model. Consequently, we approximate the objective by separating the two terms and focusing on the norm contributions:
\begin{equation}
    \left\|W_{q_1}\right\|^2_{F} \left\|(P-I)W_{k_2}^{\top}\right\|^2_{F} + \left\|W_{q_2}(P-I)^{\top}\right\|^2_{F} \left\|W_{k_1}^{\top}\right\|^2_{F}.
\end{equation}

Since $\left\|W_{q_2}\right\|_F$ and $\left\|W_{k_2}\right\|_F$ are arbitrary and can be scaled freely, we reformulate the problem to focus on the influence of $P$ relative to the protected parameters. Specifically, we consider the simplified form:
\begin{equation}
    \begin{aligned}
        \max_{P}\quad& \left\|W_q(P-I)\right\|^2_{F} + \left\|(P-I)^{\top}W_k^{\top}\right\|^2_{F} \\
        =& \left\|W_q(P-I)\right\|^2_{F} + \left\|W_k(P-I)\right\|^2_{F}.
    \end{aligned}
\end{equation}

Given the symmetry between the two terms, we can, without loss of generality, focus our analysis on a single term---e.g., $\left\|W_q(P - I)\right\|_F^2$---to simplify the optimization. By leveraging Frobenius norm and trace properties, we further reduce and analyze the problem structure, facilitating the design of effective protection mechanisms.
\begin{equation}
    \begin{aligned}
        &||W_{q}P-W_{q}||^2_{F}\\
    =&trace((W_{q}P-W_{q})(P^{\top}W_{q}^{\top}-W_{q}^{T}))\\
    =&trace(2W_{q}W_{q}^{\top}-W_{q}P^{\top}W_{q}^{\top}-W_{q}PW_{q}^{\top})
    \end{aligned}
\end{equation}

Our goal then becomes
\begin{equation}
    \begin{aligned}
        \min_P\quad &trace(W_{q}P^{\top}W_{q_1}^{\top}+W_{q}PW_{q_1}^{\top})\\
        =&trace(P^{\top}W_{q}^{\top}W_{q}+PW_{q}^{\top}W_{q_1})\\
        =&2trace(PW_{q}^{\top}W_{q})\\
        =&2trace(PU\Lambda U^{\top})\\
        =&2trace(U^{\top}PU\Lambda ).
    \end{aligned}
\end{equation}

In the above formula, we mainly use two properties of matrix trace: $trace(A^{\top})=trace(A)$ and $trace(AB)=trace(BA)$. $U\Lambda U^{\top}$ is the eigenvalue decomposition of the matrix $W_{q}W_{q}^{\top}$. Since $U$ and $P$ are both orthogonal matrices, $U^{\top}PU$ is also an orthogonal matrix with only eigenvalues 1 and -1. For a diagonal matrix $\Lambda$, the minimum of $\text{trace}(U^{\top}PU\Lambda)$ occurs at the diagonal elements of where:

\begin{itemize}
\item[$\bullet$] If $\lambda_i > 0$, then $(U^{\top}PU)_{ii} = -1$. If $\lambda_i < 0$, then $(U^{\top}PU)_{ii} = 1$.
\item[$\bullet$] If $\lambda_i = 0$, then $(U^{\top}PU)_{ii}$ can take any value.
\end{itemize}

\subsection{Proof of LWE}\label{LWE}

Because the attacker has access to the parameters of the protected model, they may attempt to recover the original model parameters by solving

\begin{equation}
    \text{diag}(\xi)W = \hat{W},
\end{equation}

where $\text{diag}(\xi)$ is a diagonal matrix containing the Taylor expansion coefficients, \(W\) represents the parameters before protection, and \(\hat{W}\) represents the parameters after protection. This effectively reduces to a matrix decomposition problem. However, this problem is highly ill-posed: the number of unknowns far exceeds the number of equations, and the system is nonlinear, making it practically unsolvable.

A more advanced adversary may resort to neural network fitting to estimate both the original parameters and the Taylor coefficients, leading to an approximate formulation:
\begin{equation}
    \text{diag}(\tilde{\xi})\tilde{W} + \epsilon = \hat{W},
\end{equation}

where \(\tilde{W}\) and \(\tilde{\xi}\) are approximations of the original parameters and Taylor coefficients, and \(\epsilon\) represents the residual error. Due to the finite precision of floating-point representation, this equation can be scaled and rounded to obtain an equivalent integer form. This transforms the problem into an instance of the Matrix Learning With Errors (Matrix-LWE) problem.

\paragraph{Hardness of Matrix Learning With Errors (Matrix-LWE).}
Matrix-LWE generalizes the standard Learning With Errors (LWE) problem. It is defined as:
\begin{equation}
    \text{diag}(\tilde{\xi})\tilde{W} + \epsilon = \hat{W} \mod q,
\end{equation}

where \(q\) denotes the maximum representable integer under floating-point arithmetic. In practice, \(q\) can be treated as sufficiently large to be ignored, reducing the problem to a standard LWE setting.

The attacker's goal is to recover the secret \(\text{diag}(\tilde{\xi})\). Matrix-LWE can be interpreted as a batch of \(\ell\) independent LWE instances:
\begin{equation}
    \begin{aligned}
        \hat{W} &= \left[ \tilde{W} \cdot \text{diag}(\tilde{\xi})_1 + \epsilon_1 \mid \tilde{W} \cdot \text{diag}(\tilde{\xi})_2 + \epsilon_2 \mid \cdots \right.\\&\left.\mid \tilde{W} \cdot \text{diag}(\tilde{\xi})_\ell + \epsilon_\ell \right] \mod q,
    \end{aligned}
\end{equation}

where each column corresponds to an LWE sample with secret \(\text{diag}(\tilde{\xi})_i\) and noise \(\epsilon_i\).

The hardness of Matrix-LWE stems from the well-established hardness of the standard LWE problem. Regev~\cite{regev2009lattices} showed that solving average-case LWE is at least as hard as solving certain worst-case lattice problems, such as the Shortest Independent Vector Problem (SIVP\(_\gamma\)) and the Gap Shortest Vector Problem (GapSVP\(_\gamma\)).

SIVP\(_\gamma\) seeks \(n\) linearly independent lattice vectors with length at most \(\gamma \cdot \lambda_n(\mathcal{L})\), while GapSVP\(_\gamma\) asks whether the shortest nonzero vector \(\lambda_1(\mathcal{L})\) in a lattice \(\mathcal{L}\) is smaller than \(d\) or greater than \(\gamma d\).

These problems are known to be NP-hard or conjectured to be intractable even for quantum algorithms. Consequently, Matrix-LWE inherits a strong worst-case-to-average-case hardness guarantee, making it a robust foundation for cryptographic primitives, particularly in post-quantum settings.

\subsection{Proof of Theorem \ref{th2}}
\label{Proof of DP}
After the model fusion, the attacker will fine-tune the fused model due to poor results. Due to the lack of data sets, the attacker may perform member stealing attacks on the model to obtain the model's data set. Considering this, the Taylor expansion mechanism may bring us natural DP properties.

Assuming that the Taylor expansion randomness of the parameters makes the noise randomness of the MLP output satisfy the Gaussian distribution $\mathcal{N}(0, \sigma^2 I)$, we can establish a parameter privacy protection mechanism, which is formally analyzed in the Appendix \ref{Proof of DP}.


\begin{theorem}\label{th2}
    Let $\phi \in (0,1)$ and $C$ be the max norm of the function $f$. If Taylor Remainder noise satisfies $\mathcal{N}(0,\sigma^2)$, where $\sigma^2>\frac{2\tilde{\Delta}^2 log(1.25/\delta)}{\epsilon^2}$, then
    for any $\delta>0$ and privacy budget $\epsilon_i$, each MLP layer $i$ output satisfies $(\epsilon_i,\delta)$-DP when $\sigma_i^2=\frac{14\phi^2NC^2\alpha}{\beta\epsilon_i}$. If we have $\alpha=\frac{\log\delta^{-1}}{\epsilon_i(1-\beta)}+1$ with $\beta\in(0,1)$ and $\sigma'^{2}=\sigma^2_i/(4C^2)$.
\end{theorem}


Here, $N$ denotes the total number of MLP layers. $C$ is the upper bound of the binary norm of the difference between the outputs of adjacent data sets.
The derived privacy guarantee ensures that each individual MLP layer satisfies $(\epsilon_i,\delta)$-DP, which has practical implications. Specifically, it indicates that the injected noise statistically obscures the model weights, making it harder for merging algorithms to align models in weight space.


\textbf{Proof}. We divide the proof of DP into two steps. In the first step, we prove that the noise on the weights makes the output of MLP satisfy the DP property. In the second step, we prove that the output with DP property will make the output of all subsequent layers satisfy the DP property.

We first present some necessary definitions and lemmas. We start by introducing differential privacy (DP), Renyi differential privacy (RDP) and $l_2$-sensitivity. 

\textbf{Definition 1} ($(\epsilon, \delta)$-Differential Privacy). A randomized mechanism $f : \mathcal{D} \to \mathcal{Y}$ satisfies $(\epsilon, \delta)$-differential privacy if for any neighbouring datasets $D, D' \in \mathcal{D}$ and $S \subset \mathcal{Y}$, it holds that $Pr[f(D) \in S] \leq e^{\epsilon} Pr [f (D') \in S] + \delta$.

\textbf{Definition 2} (Renyi differential privacy). For $\alpha>1$ and $\rho>0$, a randomized mechanism $f: \mathcal{D} \to \mathcal{Y}$ is said to have $\rho$-Renyi differential privacy of order $\alpha$ or $(\alpha, \rho)$-RDP for short, if for any neighbouring datasets $D$, $D'\in \mathcal{D}$ differing by one element, it holds that $D_{\alpha}(f(D)||f(D')):=\log\mathbb{E}(f(D)/f(D'))^{\alpha}/(\alpha-1)\leq\rho$.

\textbf{Definition 3} ($l_2$-sensitivity). The $l_2$-sensitivity $\Delta(f)$ of a function $f$ is defined as $\Delta(f)=\sup_{D,D'} ||f(D)-f(D')||_{2}$, for any two neighbouring datasets $D,D'\in\mathcal{D}$ differing by one element. 

\textbf{Step 1}. The sensitivity to the dataset $\mathcal{D}$ is defined as
\begin{equation}
    \begin{aligned}
        \tilde{\Delta}(f)&=\sup_{D,D'}||W(D)-W(D')||_2\\
        &\leq ||W(D)||_2^2+||W(D')||_2^2.
    \end{aligned}
\end{equation}

The variation of the model output (weights) in response to a single data sample is bounded. According to the Gaussian mechanism \cite{dwork2014algorithmic}, if the noise is $\mathcal{N}(0,\frac{2\tilde{\Delta}^2 log(1.25/\delta)}{\epsilon^2}\cdot I)$, after the weights are perturbed by the noise, the output of the protected model satisfies the DP property.

\textbf{Step 2}. Let us first explain the symbols. We define the output of MLP with DP property as a data set $\mathcal{D}$, and the output of all subsequent layers is defined as $\mathcal{Y}$. 

With the properly added Gaussian noises, Lemma \ref{le1} shows that the Gaussian mechanism can satisfy RDP.

\begin{lemma}\label{le1}
    For function $f:\mathcal{S}^n\to\mathcal{Y}$, the Gaussian mechanism $M = f(S) + \mathrm{u}$ with $\mathrm{u} \sim N(0,\sigma^2\mathrm{I})$ satisfies $(\alpha,\alpha \Delta^2(f))/(2\sigma^2)$-RDP. Additionally, if $M$ is applied to a subset of all the samples which are uniformly sampled from the whole datasets without replacement using sampling rate $\gamma$, then $M$ satisfies $(\alpha, 3.5\gamma^2\delta^2(q)\alpha/\sigma^2)$-RDP with $\sigma'^{2}=\sigma^2/\Delta^2(f)\geq 0.7$ and $\alpha\leq2\sigma'^{2}\log(1/\gamma\alpha(1+\sigma'^{2}))/3+1$.
\end{lemma}

We now present the following two propositions regarding RDP. Proposition \ref{pro1} shows that a composition of $N$ mechanisms satisfying RDP is also a mechanism that satisfies RDP.

\begin{proposition}\label{pro1}
    If N randomized mechanisms $f_i: \mathcal{D} \to \mathcal{Y}$ for all $i \in [N]$, satisfy $(\alpha, \rho_i)$-RDP, then their composition $(f_1(D),\dots , f_N(D))$ satisfies $(\alpha,\sum^N_{i=1} \rho_i)$-RDP.
\end{proposition}

Proposition \ref{pro2} provides the transformation from an RDP guarantee to a corresponding DP guarantee. 


\begin{proposition}\label{pro2}
    If a randomized mechanism $f: \mathcal{D} \to \mathcal{Y}$ satisfies $(\alpha, \rho)$-RDP, then $f$ satisfies $(\rho + \log(1/\delta)/(\alpha - 1), \delta)$-DP for all $\delta \in (0, 1)$.
\end{proposition}

We are now ready to give the formal proof of Theorem \ref{th2}. 
To bound the sensitivity of $y_i$, one can perform the norm clipping to restrict the $l_2$ norm of $y_i$ by replacing $y_i$ with $f_i/ \max(1, ||y_i||_2/C)$, which ensures that $||y_i||_2\leq C$. 
Due to the norm clipping of $y_i$ and the triangle inequality, the $l_2$-sensitivity of $y_i$ could be bounded as
\begin{equation}\label{51eq}
    \sup_{D,D'}||y_i(D)-y_i(D')||_2\leq 2C,
\end{equation}

where $D,D'$ are any two neighbouring datasets differing by one element. By Eq (\ref{51eq}) and Lemma \ref{le1}, the noised output satisfies $(\alpha, 14\phi^2C^2\alpha/\sigma^2)$-RDP. Substituting $\sigma_i^2=\frac{14\gamma_2\phi^2NC^2\alpha}{\beta\epsilon_i}=\frac{14\gamma_2\phi^2NC^2\alpha}{\epsilon_i+\frac{\log\delta}{\alpha-1}}$ shows that $M_i$ satisfies $(\alpha,\epsilon_i+\frac{\log\delta}{\alpha-1})$-RDP. Proposition \ref{pro2} show that $y_i$ satisfies $(\epsilon_i, \delta)$-DP. Then applying Proposition \ref{pro1} again shows that the composite message mechanism $Y_i = \{y^t_i\}^{T}_{t=1}$ is $(\alpha, \epsilon_i+\frac{\log\delta}{\alpha-1})$-RDP. Lastly, the proof is completed by applying Proposition \ref{pro2} to translate $(\alpha, \epsilon_i+\frac{\log\delta}{\alpha-1})$-RDP to $(\epsilon_i, \delta)$-DP for mechanism $y_i$, which concludes the proof.

\subsection{High-order Derivate of $GELU(\cdot)$}\label{gelu}

The $GELU(\cdot)$ activation function is the standard Gaussian cumulative distribution function. According to the Leibniz rule, we have
\begin{equation}
    \begin{aligned}
        gelu^{(n)}(x)&=\sum_{k=0}^{n}\begin{pmatrix}
        k\\
        n
    \end{pmatrix}x^{(k)}\Phi^{(n-1)}(x)\\
    &=x\Phi^{(n)}(x)+n\Phi^{(n-1)}(x).
    \end{aligned}
\end{equation}

We let 
\begin{equation}\label{recursive}
    \begin{aligned}
        h_n(x)&=\frac{1}{\sqrt{2\pi}}(e^{-\frac{x^2}{2}})^{(n)}\\
        &=(-x)\frac{1}{\sqrt{2\pi}}(e^{-\frac{x^2}{2}})^{(n-1)}-(n-1)\frac{1}{\sqrt{2\pi}}(e^{-\frac{x^2}{2}})^{(n-2)}\\
        &=(-x)h_{n-1}-(n-1)h_{n-2}.
    \end{aligned}
\end{equation}

The n-order derivative of $\Phi^{(n)}$ is given b
\begin{equation}
    \Phi^{(n)}(x)=\frac{1}{\sqrt{2\pi}}(e^{-\frac{x^2}{2}})^{(n-1)}=h_{n-1}(x).
\end{equation}

The n-order derivative of $GELU(\cdot)(x)$ is given by 
\begin{equation}
    gelu^{(n)}(x)=xh_{n-1}(x)+nh_{n-2}(x),
\end{equation}

where $h_{n}(x)$ can be recursively computed by Eq. (\ref{recursive}) and $h_0(x)=\frac{1}{\sqrt{2\pi}}e^{-\frac{x^2}{2}}$. 

\subsection{High-order Derivate of $SiLU(\cdot)$}\label{silu}

The $SiLU(\cdot)$ activation function is given by 
\begin{equation}
    \begin{aligned}
            SiLU^{(n)}(x)&=-\sum_{k=0}^{n}\begin{pmatrix}
        k\\
        n
    \end{pmatrix}x^{(k)}\sigma^{(n-k)}(x)\\
    &=x\sigma^{(n)}(x)+n\sigma^{(n-1)}(x).
    \end{aligned}
\end{equation}

We let
\begin{equation}
    \begin{aligned}
        h_n(x)&=\sigma^{(n)}(x)\\
        &=[\sigma(x)(1-\sigma(x))]^{(n-1)}\\
        &=\sum_{k=0}^{n-1}\begin{pmatrix}
            k\\
            n-1
        \end{pmatrix}\sigma^{(k)}(x)(1-\sigma(x))^{(n-k)}\\
        &=\sum_{k=0}^{n-1}\begin{pmatrix}
            k\\
            n-1
        \end{pmatrix}\sigma^{(k)}(x)(-\sigma(x))^{(n-k)}\\
        &=-\sum_{k=0}^{n-1}\begin{pmatrix}
            k\\
            n-1
        \end{pmatrix}h_{k}(x)h_{n-k-1}(x).\\
    \end{aligned}
\end{equation}

The n-order derivative of $SiLU(\cdot)$ is given by
\begin{equation}
    SiLU^{(n)}(x)=xh_n(x)+nh_{n-1}(x),
\end{equation}

where $h_n(x)$ can be recursively computed by Eq. () and $h_0(x)=\sigma(x)$.

\section{Accelation of Eigenvalue Decomposition}\label{Accelation}
As shown in Theorem~\ref{th1}, computing the optimal orthogonal matrix \(P\) requires the eigen-decomposition of \(W_q^\top W_q\), typically via SVD. However, exact SVD is impractical for high-dimensional \(W_q\) due to its \(O(n^3)\) cost. To address this, we use Randomized SVD (RSVD), which exploits the low-rank structure of weight matrices for efficient decomposition in a reduced subspace.

\textbf{Step 1: Random Projection.} Generate a random matrix $\Omega \in\mathbb{R}^{n\times k}$ from a standard normal distribution, where $k\ll n$ controls the trade-off between accuracy and efficiency. Compute $Y=S\Omega$, obtaining an $n\times k$ projection of $S$. 

\textbf{Step 2: Low-Dimensional SVD.} Perform QR decomposition on $Y$, yielding an orthogonal matrix $Q$. Compute the reduced matrix $C=Q^{\top}SQ$ and apply SVD: $C=U_C\Sigma V_C^{\top}$, where $U_C, V_C\in \mathbb{R}^{k\times k}$ are orthogonal, and $\Sigma$ is diagonal.

\textbf{Step 3: Reconstruction.} Map singular vectors back via $U=QU_C, V=QV_C$, yielding an approximation $S\approx U\Sigma V^{\top}$. 

Random projection, QR decomposition and SVD on $k\times k$ matrix has $O(n^2k)$, $O(nk^2)$ and $O(k^3)$ computational complexity respectively. The total computational complexity is $O(n^2k+nk^2+k^3)$, significantly faster than $O(n^3)$, making RSVD ideal for large-scale models.

\section{Appendix / Decode PaRaMS}\label{Attacking PaRaMS}

In this section, we analyze the security of the PaRaMS protection mechanism under an adversary with knowledge of the base model. We first summarize the defense strategies of PaRaMS (Sec. F.1) and then present the attack pipeline (Sec. F.2).

\subsection{PaRaMS Defense Mechanism}
PaRaMS protects model weights using two main techniques:

\begin{enumerate}
\item \textbf{MLP Parameter Rearrangement}: The defender applies a random permutation to the weight matrices $W_1$ and $W_2$ of an MLP, using a permutation matrix $P$ such that:
\begin{equation}
W_1' = P W_1, \quad W_2' = W_2 P^T
\end{equation}
This transformation ensures functional equivalence while disrupting parameter alignment for model merging.

\item \textbf{Random Multi-head Scaling}: In the multi-head attention block, each head $i$ is assigned two diagonal scaling matrices $A_i$ and $B_i$, drawn from a uniform distribution $U(s_{min}, s_{max})$, modifying the parameters as follows:
\begin{equation}
    \begin{aligned}
        &Q_i \leftarrow A_i Q_i, \quad K_i \leftarrow A_i^{-1} K_i, \quad V_i \leftarrow B_i V_i, \\
        &\quad W_O[:, i] \leftarrow W_O[:, i] B_i^{-1}
    \end{aligned}
\end{equation}
These transformations keep the attention output functionally equivalent while altering parameter values.
\end{enumerate}

\subsection{Attack Pipeline}
We now describe our attack methodology for recovering the protected parameters of both MLP and attention layers. The key insight is that PaRaMS transformations preserve functional equivalence but fail when the base model is known.

\subsubsection*{MLP Recovery.}
The MLP protection strategy involves shuffling weight matrices using a permutation matrix:

\begin{equation}
W_1' = P W_1, \quad W_2' = W_2 P^T.
\end{equation}

We break down the recovery process into two steps:

\begin{enumerate}
\item \textbf{Recovering the Permutation Matrix}:
\begin{itemize}
\item Since it is transformed along rows or columns, we analyze column-wise correspondences between the protected model and the base model.
\item Specifically, we compute the distance between each column vector in and those in. The columns in  are matched to those in  based on the smallest  distance:
\begin{equation}
P(i) = \arg\min_j | W_1'[:, i] - W_1[:, j] |_2.
\end{equation}
\item By iterating over all columns, we recover the permutation order.
\end{itemize}

\item \textbf{Undoing the Permutation}:
\begin{itemize}
\item Once the correct mapping is found, we apply the inverse permutation  to restore :
\begin{equation}
\hat{W_1} = P^{-1} W_1'.
\end{equation}
\item With  recovered, we use  to reverse the transformation on :
\begin{equation}
\hat{W_2} = W_2' P.
\end{equation}
\end{itemize}
\end{enumerate}

By following these two steps, we effectively undo the obfuscation and recover the original MLP parameters.

\subsubsection*{Attention Scaling Recovery.}
Our approach follows a variance-based differentiation method similar to prior attacks on filter permutation defenses.

\begin{enumerate}
\item \textbf{Estimating the Scaling Factors}:
\begin{itemize}
\item Since the protection scheme scales the query, key, and value matrices with diagonal matrices \( A \) and \( B \), we estimate these scaling factors by computing the element-wise division between the protected and base model’s parameters:
\begin{equation}
A_{ii} \approx \frac{Q'_{ii}}{Q_{ii}}, \quad A_{ii} \approx \frac{K'_{ii}}{K_{ii}}, \quad B_{ii} \approx \frac{V'_{ii}}{V_{ii}}.
\end{equation}
\item As \( A \) and \( B \) are diagonal, we obtain their estimated values using the median scaling across the diagonal elements.
\end{itemize}

\item \textbf{Recovering the Original Parameters}:
\begin{itemize}
    \item Using the estimated \( A \) and \( B \), we apply their inverse transformations to recover the original weights:
    \begin{equation}
        \hat{Q} = A^{-1} Q', \quad \hat{K} = A K'
    \end{equation}
    \item Similarly, for the value matrix and output projection:
    \begin{equation}
        \hat{V} = B^{-1} V', \quad \hat{W_O} = W_O' B.
    \end{equation}
\end{itemize}
\end{enumerate}

\section{Appendix / Other Related Work}\label{Other Related Work}

\highlight{Watermarking.} 
Watermarking is a widely used technique for ownership verification, embedding identifiable patterns into model weights. However, existing watermarking methods fail to remain effective in model merging. Since watermarks rely on specific weight quantization properties, the merging process disrupts these properties, rendering the watermark unrecognizable.

\highlight{Fingerprinting.}
Fingerprinting ties ownership verification to model behavior rather than weights, making it resistant to merging. However, as a passive protection method, fingerprinting cannot prevent unauthorized merging or usage. Since it does not alter model functionality, attackers can still merge models freely and remove fingerprints using adversarial techniques. In contrast, MergeBarrier is an active protection mechanism that deters unauthorized merging through intrinsic model constraints.

\highlight{Authorization-based Protection.} 
Authorization-based approaches, such as key-based and TEE-based inference, effectively prevent merging by limiting model functionality in unauthorized scenarios. However, these methods require external components (e.g., TEEs or cryptographic keys), which interfere with the open-source nature of model sharing and limit their practicality in typical open-source settings. In contrast, our method ensures protection without introducing external dependencies, preserving it practical for open-source models.

\section{Appendix / Sharpness vs. Accuracy}\label{sharpness}

In deep learning, \textbf{sharpness} characterizes how sensitive the loss function is to small perturbations in model parameters. Formally, for a model with parameters $w$ and loss function $\mathcal{L}(w)$, the $\epsilon$-sharpness is defined as:
\begin{equation}
    \text{Sharpness}_\epsilon(w) = \max_{\|\delta\| \leq \epsilon} \mathcal{L}(w + \delta) - \mathcal{L}(w)
\end{equation}

A commonly used smoothed version normalizes by the scale of the original loss:
\begin{equation}
\text{Sharpness}_\epsilon(w) = \frac{\mathcal{L}(w + \delta) - \mathcal{L}(w)}{1 + \mathcal{L}(w)}, \quad \text{s.t. } \|\delta\| \leq \epsilon
\end{equation}

Here, $\delta$ represents a small perturbation in the parameter space. A \emph{sharp minimum} corresponds to large loss increases with small $\delta$, while a \emph{flat minimum} corresponds to low sensitivity.

Sharpness is fundamentally a geometric measure of the curvature of the loss landscape around a local minimum. It provides insight into the model's robustness to parameter perturbations and is empirically linked to generalization. In particular:

\begin{itemize}[leftmargin=1.5em]
    \item Models converging to \textbf{flat minima} tend to generalize better, since they are more stable under small changes in weights.
    \item \textbf{Sharp minima} suggest the model may have overfit the training data, being sensitive to minor variations and less robust under distribution shifts.
    \item Optimizers such as SGD implicitly favor flatter minima compared to adaptive methods.
    \item Regularization techniques (e.g., weight decay or adversarial training) can reduce sharpness, enhancing generalization.
\end{itemize}

Sharpness-aware training algorithms (e.g., SAM -- Sharpness-Aware Minimization) explicitly incorporate this metric to improve generalization by minimizing both the loss and local sharpness.

\begin{itemize}[leftmargin=1.5em]
    \item \textbf{Definition}: Accuracy is a task-dependent performance metric measuring the proportion of correct predictions. Sharpness, in contrast, is a task-independent property of the loss landscape.
    \item \textbf{Scope}: Accuracy evaluates the model's performance on data. Sharpness evaluates the stability of the model's solution in weight space.
    \item \textbf{Granularity}: Accuracy is a discrete, outcome-based metric. Sharpness is continuous and sensitive to geometric structure.
    \item \textbf{Generalization}: Sharpness provides predictive value for generalization capability, particularly when training and validation performance are similar. Accuracy alone cannot distinguish overfitting from well-generalized performance.
    \item \textbf{Optimization}: Accuracy is typically optimized directly during training. Sharpness is rarely optimized directly unless specialized objectives (e.g., SAM) are used.
\end{itemize}

Accuracy and sharpness are complementary. While accuracy captures task-specific correctness, sharpness reveals the robustness and generalization behavior of the learned solution. An ideal evaluation pipeline considers both: high accuracy to ensure performance, and low sharpness to ensure stability and reliability.

\end{document}